\documentclass[10pt,journal,compsoc]{IEEEtran}
\usepackage{comment}
\usepackage[italian, main=english]{babel}
\usepackage{caption}
\usepackage{adjustbox}
\usepackage{algorithm}
\usepackage{graphicx}
\usepackage{listings}
\usepackage{hyperref}
\usepackage{multirow}
\usepackage{multicol}
\usepackage{setspace}
\usepackage{tabularx}
\usepackage{color}
\usepackage{float}
\usepackage{wrapfig}
\usepackage{booktabs}
\usepackage{array}
\usepackage{vcell}
\usepackage{changepage}
\usepackage{subfig}

\usepackage{tikz}
\usetikzlibrary{shapes,arrows,positioning}
\usepackage{pgfplots}
\pgfplotsset{compat=newest}

\usepackage{amsmath,amssymb,amsfonts}
\usepackage[flushmargin]{footmisc}
\usepackage{listings}
\usepackage{bm}
\usepackage[inline]{enumitem}
\usepackage{soul}
\usepackage{algpseudocode}
\usepackage{algorithm}
\usepackage{adjustbox}
\usepackage{fancybox}
\algnewcommand\algorithmicforeach{\textbf{for each}}
\algdef{S}[FOR]{ForEach}[1]{\algorithmicforeach\ #1\ \algorithmicdo}

\newcolumntype{P}[1]{>{\centering\arraybackslash}p{#1}}

\newboolean{showcomments}
\setboolean{showcomments}{true}         

\ifthenelse{\boolean{showcomments}}
  {\newcommand{\nb}[2]{
  \fbox{\bfseries\sffamily\scriptsize#1}
     {\sf\small$\blacktriangleright$\textit{\textcolor{red}{#2}}$\blacktriangleleft$}
   }
  }
  {\newcommand{\nb}[2]{}
   
  }

\newcommand{\methodname}{MobHide}
\newcommand{\toolname}{HideDroid}

\definecolor{codegreen}{rgb}{0,0.6,0}
\definecolor{codegray}{rgb}{0.5,0.5,0.5}
\definecolor{codepurple}{rgb}{0.58,0,0.82}
\definecolor{backcolour}{rgb}{0.95,0.95,0.92}

\lstdefinestyle{mystyle}{
    backgroundcolor=\color{backcolour},   
    commentstyle=\color{codegreen},
    keywordstyle=\color{magenta},
    numberstyle=\scriptsize\color{codegray},
    stringstyle=\color{codepurple},
    basicstyle=\ttfamily\scriptsize,
    breakatwhitespace=false,         
    breaklines=true,                 
    captionpos=b,                    
    keepspaces=true,                 
    numbers=left,
    numbersep=-7pt,                  
    showspaces=false,                
    showstringspaces=false,
    showtabs=false,                  
    tabsize=2,
}

\ifCLASSOPTIONcompsoc
  \usepackage[nocompress]{cite}
\else
  \usepackage{cite}
\fi
\ifCLASSINFOpdf
\else
\fi
\begin{document}

\title{You Can’t Always Get What You Want: Towards User-Controlled Privacy on Android}


\author{Davide Caputo,
        Francesco Pagano,
        Giovanni Bottino,
        Luca Verderame,
        and~Alessio~Merlo
        
        
\thanks{Corresponding Author: Alessio Merlo.}

\IEEEcompsocitemizethanks{

\IEEEcompsocthanksitem D. Caputo is with the University of Genova, Genova, 16146, Italy. 
\protect\\
Email: davide.caputo@dibris.unige.it

\IEEEcompsocthanksitem F. Pagano is with the University of Genova, Genova, 16146, Italy.
\protect\\
Email: francesco.pagano@dibris.unige.it 

\IEEEcompsocthanksitem G. Bottino is with the University of Genova, Genova, 16146, Italy.
\protect\\
Email: giovannibottino3@gmail.com

\IEEEcompsocthanksitem L. Verderame is with the University of Genova, Genova, 16146, Italy.
\protect\\
Email: luca.verderame@dibris.unige.it

\IEEEcompsocthanksitem A. Merlo is with the University of Genova, Genova, 16146, Italy.
\protect\\
Email: alessio.merlo@dibris.unige.it

}
}

\markboth{IEEE Transactions on dependable and Secure Computing,~Vol.~X, No.~Y, Month~yyyy}%
{Shell \MakeLowercase{\textit{et al.}}: Bare Demo of IEEEtran.cls for Computer Society Journals}


\IEEEtitleabstractindextext{%
\begin{abstract}

Mobile applications (hereafter, apps) collect a plethora of information regarding the user behavior and his device through third-party analytics libraries. However, the collection and usage of such data raised several privacy concerns, mainly because the end-user - i.e., the actual owner of the data - is out of the loop in this collection process. Also, the existing privacy-enhanced solutions that emerged in the last years follow an "all or nothing" approach, leaving the user the sole option to accept or completely deny access to privacy-related data.

This work has the two-fold objective of assessing the privacy impact of mobile analytics libraries and proposing a data anonymization methodology that offers a trade-off between the utility and privacy of the collected data and enables complete control over the sharing process.
To achieve that, we present an empirical privacy assessment on the analytics libraries used in the $4500$ most-used Android apps of the Google Play Store in late 2020. 
Then, we propose an empowered anonymization methodology, based on MobHide \cite{mobhide}, that gives the end-user complete control over the collection and anonymization process.
Finally, we empirically demonstrate the applicability and effectiveness of our solution thanks to HideDroid, a fully-fledged anonymization app for the Android ecosystem.

\end{abstract}

}

\maketitle

\IEEEdisplaynontitleabstractindextext

%
\IEEEpeerreviewmaketitle

\IEEEraisesectionheading{\section{Introduction}\label{sec:introduction}}

\IEEEPARstart{T}{he} high number of mobile apps currently available for Android (nearly $4.83M$ in mid-2020 \cite{num_app_available}) forced developers and companies to increase the quality of their apps in order to emerge in a fiercely competing market.

Users tend to choose the app to install according to both the number of its features and the ratings provided by customers  \cite{appbrain_analytics,increase_app_downloads_appradar}. 

Thus, apps aim to maximize the user experience and tailor content to satisfy the users' expectations.
Such a process forced developers to collect data about the users and their interaction with the apps, in order to evaluate their behavior and preferences, and enhance the app accordingly \cite{privacy_risk_analysis}.

To this aim, analytics libraries allow developers to collect, filter, and analyze those data programmatically. 
They are typically composed of two elements: a client (i.e., an SDK), included in the app, and a back-end service.
The SDK is responsible for collecting information regarding the device, the user, and her interaction with the app. The collected data is packed in data structures called \emph{events} and sent to the back-end service through the network.
The back-end, hosted as a cloud service, aggregates the received events and provides a dashboard for developers to analyze and filter the data.

The ease of use of analytics libraries, such as \textit{Facebook Analytics} and \textit{Google Firebase Analytics}, attracted a wide range of developers, leading to their rapid and widespread diffusion in the most popular mobile apps\cite{Exodus_analytics_library}.
However, the adoption of such libraries raised several concerns on the privacy of the collected information, as described in \cite{dynamic_leakage_analysis_android} and \cite{privacy_risk_analysis}.
For instance, several works (e.g., \cite{information_leakage_through_mobile_analytics_services,user_privacy_in_android_libraries}) reported how analytic libraries share the same privileges and resources of the hosting app and are able to access and collect sensitive information regarding the users and their behaviors without proper privacy-preserving mechanisms.
Furthermore, authors in \cite{Verderame3PDroid} demonstrated that only a negligible part of apps fulfills the Google Play privacy requirements for Android apps ($1\%$ out of the $5473$ most downloaded apps in 2020).

The privacy issues of analytics libraries attracted the research community, which proposed several solutions to enhance the privacy of the collected data through anonymization techniques. 
For instance, Zhang et al. \cite{frequency_profiling} proposed a solution allowing the developer to anonymize the collected information according to differential privacy techniques, while Liu et al. \cite{privacy_risk_analysis} designed an Android app able to intercept and block all the API related to analytics libraries.
Also, Razaghpanah et al. \cite{tracking_app_analysis} developed an app to block the network requests containing personal information.

Unfortunately, state-of-the-art solutions have some limitations.
First, they do not give any control on the collected data and the anonymization process to the end-user, which is the actual data owner. 
In such a scenario, the anonymization solutions either autonomously select the type of data to collect and anonymize, or leave this choice to the app developer.
Also, the proposed anonymization solutions follow an ``all or nothing" approach, giving the sole option to either fully accept or deny the collection of personal data. Thus, the app developer can access the complete set of non-anonymized data (100\% utility of data, 0\% privacy) or cannot access any information at all (0\% utility, 100\% privacy).
Finally, existing solutions require invasive technical requirements such as adopting a customized OS, executing on a rooted device, the prior knowledge of personal data, or they require to customize the app logic. As a consequence, they could hardly be adopted in the wild.

\textbf{Contributions of the paper.} In this work, we seek to address the following research questions (RQs): 
\begin{itemize}
    \item \textbf{RQ1}: 
    \textit{How widespread is the adoption of analytics libraries in mobile apps? Which are the most used libraries?}

    \item \textbf{RQ2}: 
    \textit{What is the impact of analytics libraries on the overall network traffic generated by the apps?}

    \item \textbf{RQ3}: \textit{Which pieces of information are collected by analytics libraries, and how are they relevant for the users' privacy?}
   
    \item \textbf{RQ4}: \textit{Is it possible to apply a local anonymization strategy compatible with existing analytics libraries that may grant the user a fine-grained control over the privacy of her data?}
    
\end{itemize}

We conducted an extensive experimental campaign over the first $4500$ most downloaded Android apps from the Google Play Store between November 2020 and January 2021. We analyzed each app both statically and dynamically to evaluate the impact on privacy caused by the use of analytics libraries. 
Furthermore, we classified each collected data using the concepts of \emph{Explicit Identifiers}, \emph{Quasi Identifiers}, and \emph{Sensitive Data} \cite{DataPrivacyFusion2012} to evaluate the privacy impact of the data collection process. 
Then, we empowered the \methodname{} methodology, proposed in our previous work \cite{mobhide}, and we developed \toolname{} (publicly available on GitHub \cite{HideDroidGithub}), a full-fledged anonymization app for Android. Finally, we tested \toolname{} against the $4500$ apps of the experimental testbed to assess the efficacy and applicability of local anonymization strategies according to the user's preferences.
As an additional contribution, we released the entire dataset of anonymized network requests generated during the experimental campaign \cite{dataset}.

\textit{\textbf{Structure of the paper.}} The rest of the paper is organized as follows: Section \ref{sec:background} introduces the functionalities of analytics libraries and some basic concepts on data anonymization, while Section \ref{sec:analytics-in-the-wild} presents the in-the-wild privacy analysis of the usage of analytics libraries in mobile apps. Section \ref{sec:mobhide-hidedroid} details the \methodname{} methodology, describes the \toolname{} anonymization app, and presents the evaluation of the usability and effectiveness of our solution. 
Section \ref{sec:related-work} discusses the current state-of-the-art, while Section \ref{sec:conclusion} concludes the paper and points out some future extensions of this work.

\section{Background}\label{sec:background}
\subsection{Analytics Libraries}\label{sec:analytics-libraries}

Analytics libraries are software solutions that allow developers to monitor and track the user's interactions with their apps. They are typically composed of two parts, namely the client library and the back-end system. 
The client library consists of a Software Developer Kit (SDK), containing a set of APIs for the in-app data collection and the auxiliary scripts to include and compile the client inside the app package. 
The back-end system, either available as a cloud service or an on-premise solution, is responsible for collecting and aggregating the clients' data and gives the developer a full-featured dashboard to review, analyze, and extract the requested information.

Analytics libraries enable collecting a wide range of information belonging to two macro-categories: \emph{personal data} and \emph{event data} \cite{facebook_event,firebase_analytics,amplitude_events}.
Personal data include details about the user, such as the email address or the username, and the device, e.g., device name, SDK version, and the network carrier.
Event data focus on the interactions between the user and the app. Analytics libraries provide to developers \emph{i)} a set of predefined events, like \textit{``app open"} or \textit{``app close"}, and \emph{ii)}  the possibility to create custom events. Examples of custom events include: \textit{``add to cart"}, \textit{``add payment info"}, and \textit{``purchase"} in case of e-commerce apps, or \textit{``click to Ad"}, \textit{"rewarded video"} in case of mobile games.

Personal and event data are encoded in key-value data structures and sent by the client library to the back-end system using the network connectivity, e.g., through HTTPS connections.

\subsection{Anonymization Techniques}\label{sec:anonymization}

In data privacy, the set of attributes in a microdata set \cite{DataPrivacyFusion2012,di2011anonymization} can be mainly divided into three categories:
\begin{itemize} 
    \item \textit{Explicit Identifiers (EI)}. EI are user-identifying attributes, such as the name/surname, the social security number (SSN), or the Insurance ID.
    \item \textit{Quasi-Identifiers (QI)}. This category includes attributes that can be combined with other external data sources (e.g., publicly available databases) to indirectly identify a user.
    Examples of QI include geographic and demographic information, phone numbers, and e-mail IDs. 
    \item \textit{Sensitive Data (SD)}. SD are attributes that contain relevant information for the recipient of the microdata set, like, e.g., health diseases, salaries, eating habits, just to cite a few.
\end{itemize}

\noindent
Data Anonymization (DA) techniques aim to decouple the user's identity (i.e., EI and QI) from her sensitive information (i.e., SD) before releasing the microdata set in the wild. To do so, DA techniques first remove or substitute the EI and then alter the QI set to reduce the possibility to re-identify the user through external databases, and then correlate her identity with the corresponding SD attributes.

DA techniques can be divided into two groups: \textit{Perturbative (PT)} and \textit{Not Perturbative (NPT)}. 
\textit{PT} techniques consist of altering the QI data with dummy information to weaken their correlation.
For instance, a numeric attribute, e.g., the \textit{zip\_code = $16011$}, can be transformed to \textit{zip\_code = $16129$} by adding a noise equal to $118$. 
\textit{NPT}  techniques aim at reducing the detail in the data through generalization of values, with a very limited impact on the semantics of data. As an example, the value \textit{zip\_code = $16011$} can be generalized to \textit{zip\_code = $160**$}.

The application of each anonymization technique can be evaluated according to the level of \textit{privacy} and \textit{utility} of the processed data. The two values are inversely proportional: the more anonymization is applied, the higher level of user's privacy is granted (i.e., the probability to de-anonymize the user is low), at the expense of the utility of the data (i.e., the semantics data is highly affected). Conversely, if the level of anonymization is low,  the utility of data is high, but the level of user's privacy is reduced, thus raising the probability to de-anonymize a user in the released microdata set \cite{li2009tradeoff,di2011anonymization}.

Unfortunately, PT techniques have several limitations in terms of utility and privacy. Indeed, complex noise transformations severely alter the semantics of data resulting in a significant utility loss, as described in \cite{DataPrivacyFusion2012,li2009tradeoff}. 
On the other hand, simple noise distortion techniques can be reverted to obtain the original microdata set, as demonstrated by \cite{DataPrivacyFusion2012,kifer2006injecting,mivule2013utilizing}.

In our work, we focused mainly on two NPT anonymization techniques: \textbf{Data Generalization} (DG) \cite{loukides2012utility} and \textbf{Differential Privacy} (DP)\cite{di2011anonymization}.

\textbf{Data Generalization.} DG replaces specific values of a set of attributes belonging to the same domain, with more generic ones \cite{samarati2001protecting}. 
In a nutshell, given an attribute \textit{$A$} belonging to a domain \textit{$D_0(A)$}, it is possible to define a \textbf{Domain Generalization Hierarchy (DGH)} for a Domain (\textit{$D$}), as a set of \textit{n} anonymization functions \textit{$f_h$}: $h = 0$, \dots, $n-1$, such that:
    \begin{equation}
        \textit{$D_0$} \xrightarrow{\textit{$f_0$}} 
        \textit{$D_1$} \xrightarrow{\textit{$f_1$}}
        \dots \xrightarrow{\textit{$f_n-1$}} \textit{$D_n$}, 
    \end{equation}
and:   
    \begin{equation}
    \textit{$D_0({A})$} \subseteq \textit{$D_1({A})$} \subseteq \dots \subseteq \textit{$D_n({A})$}
    \end{equation}

It is worth noticing that the more generalization functions are invoked on the original data, the higher is the resulting privacy value (and the lower is the data utility), as heterogeneous data are transformed into a more reduced set of general values.
Generalization techniques are suitable for semantically independent data, such as the set of \emph{personal data} collected by analytics libraries (e.g., the \textit{deviceName} or the \textit{phone number}).

\textbf{Differential Privacy.} The DP technique consists of altering the original distribution of a set of interdependent data using a perturbation function \cite{dwork2008differential}.
This approach is usually applied in a context where i) the main requirement is the confidentiality of the data exchanged between pairs, and ii) the receivers' identity is unknown a priori.
There are two main models for defining DP problems: \textit{centralized} and \textit{local} model. 
In the centralized model, the data are sent to a trusted entity that applies DP algorithms and then shares the anonymized dataset with an untrusted third-party client \cite{dwork2008differential}.
On the contrary, the local model assumes all external entities and communication channels as untrusted \cite{cormode2018privacy,yang2020local}.
In such a situation, local DP techniques aim at performing the data perturbation locally before releasing any dataset to an external party.

In our scenario, we consider the user as the sole owner of its data, and we trust neither the analytics company nor the developer. 
To this aim, the local DP model is suitable to anonymize sequences of events logged by analytics libraries. The objective of DP is to transform a sequence of events \textit{($e_1, e_2, \dots, e_n$) $\in D$ } in a different sequence of events \textit{($z_1, z_2, \dots, z_n$) $\in D$} through the application of a perturbation function $R: D \rightarrow D$ to each event. 
This function is commonly a probability distribution, defined a priori: \textit{$z_i = R(e_i)$}.

\section{Analytics Libraries In The Wild}\label{sec:analytics-in-the-wild}

In this section, we evaluate the presence of mobile analytics libraries in Android apps and their impact on the security of the device and the user (RQ1-RQ3) by conducting an experimental campaign on the $4500$ most downloaded Android apps taken from Google Play Store between November 2020 and January 2021.

Despite there exist some other works that analyze the spread of third-party libraries on mobile apps (e.g.,\cite{zhangprivaid,privacy_risk_analysis}), such proposals have some important limitations: i) they investigate the network traffic of third-party libraries for specific categories of apps only (e.g., parental control apps \cite{feal2020angel}, paid apps \cite{han2020price} or pre-installed apps \cite{gamba2020analysis}), ii) the analysis does not inspect the content of network requests and its privacy impacts, and iii) they do not share a public dataset that could be used for future research activities.
Those considerations drove us to present a new analysis with a specific focus on mobile analytics libraries that allowed for the creation of an updated dataset that we publicly released to the research community \cite{dataset}.

To address \textbf{RQ1}, we statically analyzed all the downloaded apps to identify the presence of analytics libraries in the app code. 
Each Android app has been scanned using \textit{Androguard} \cite{Androguard} and \textit{Exodus Core} \cite{Exodus_core} to detect if it includes package names belonging to analytics libraries (e.g., the package \textit{com.google.firebase.crashlytics} refers to the use of Google CrashLytics library).

Table \ref{tab:analytics_services} summarizes the most included libraries in the dataset. 
It is worth noticing that the most used analytics libraries belong to Google (92.5\%) and Facebook (52.8\%). 

\begin{table}[t]
    \scriptsize
    \centering
    \begin{tabular}{||c|c|c||}
         \hline
    \textbf{Analytics Library} & \textbf{\# App} & \textbf{Percentage} \\
        \hline
    Google Firebase Analytics & $3569$ & $79.3\%$ \\
        \hline
    Google AdMob & $3371$ & $74.9\%$ \\
        \hline
    Google CrashLytics & $2093$ & $46.5\%$ \\
        \hline
    Facebook Login & $1688$ & $37.5\%$ \\
        \hline
    Facebook Ads & $1616$ & $35.9\%$\\
        \hline
    Facebook Share & $1580$ & $35.1\%$\\
        \hline
    Facebook Analytics & $1517$ & $33.7\%$ \\
        \hline
    Unity 3d Ads & $1244$ & $27.2\%$\\
        \hline
    Google Analytics & $1103$ & $24.5\%$ \\
        \hline
    Moat & $1008$ & $22.4\%$ \\
        \hline
    Google Tag Manager & $873$ & $19.4\%$ \\
        \hline
    AppLovin & $842$ & $18.7\%$ \\
        \hline
    Facebook Places & $806$ & $17.9\%$\\
        \hline
    ironSource & $801$ & $17.8\%$ \\
        \hline
    \end{tabular}
    \caption{Distribution of analytics libraries in the top $4500$ Android apps.}
    \label{tab:analytics_services}
\end{table}

\noindent
The analysis of the impact of analytics libraries on the network traffic generated by apps (\textbf{RQ2}) as well as on the privacy of the user and the device (\textbf{RQ3}) required a dynamic analysis phase to evaluate the behavior of the different analytics solutions \emph{at runtime}. Also, dynamic analysis allowed detecting also obfuscated or runtime-loaded libraries that can hardly be detected through static analysis.

To do so, we tested each app for $10$ minutes to collect the generated network traffic. Apps were tested using DroidBot \cite{droidbot}, a black-box framework that automatically stimulates the app under test (AUT) by mimicking human interactions. 
The testing environment consists of an emulated Android device with Android OS version 10 and root permissions. 
For the traffic collection, we relied on \textit{mitmproxy} \cite{mitmproxy}. 
This tool enables the deep inspection of SSL connections thanks to the installation of a custom CA certificate in the system certificate directory (allowed by root permissions). 
To further cope with apps and libraries implementing SSL Pinning techniques \cite{android_https,towardshttps} to protect the network traffic, we dynamically instrumented each AUT using Frida \cite{FridaAndroid} in order to bypass the most common implementations of SSL pinning.

The dynamic analysis has been carried out in an Ubuntu 20.04 VM with 32GB of RAM and 16 cores at 3.8 GHz.
To speed up the evaluation phase, we used three Android emulator instances at a time. 
The analysis lasted 14 days and collected $265770$ unique network requests generated by the AUTs.

The collected traffic has been inspected to determine if it belongs to an analytics library. 
In detail, we classified all the network traffic according to a list of well-known network hosts connected to analytics libraries (extracted through the Exodus tool \cite{Exodus_core}). 

Moreover, we proposed a heuristic based on the parsing of the network requests according to a set of keywords (e.g., \textit{device-name}, \textit{device-id}, \textit{device-info}, \textit{event-type}, \textit{event-info}, \textit{event} or \textit{event-name}), to cope with the possible presence of unknown hosts. Such keywords are typically included in events generated by analytics libraries \cite{mobhide}. If the request contains at least one of them, the heuristic labels the request and keeps track of the new host.
Finally, any host identified using the heuristic has been manually inspected to detect and remove false positives.

\begin{figure}[t]
\centering
    \subfloat[]{\includegraphics[width=0.5\linewidth]{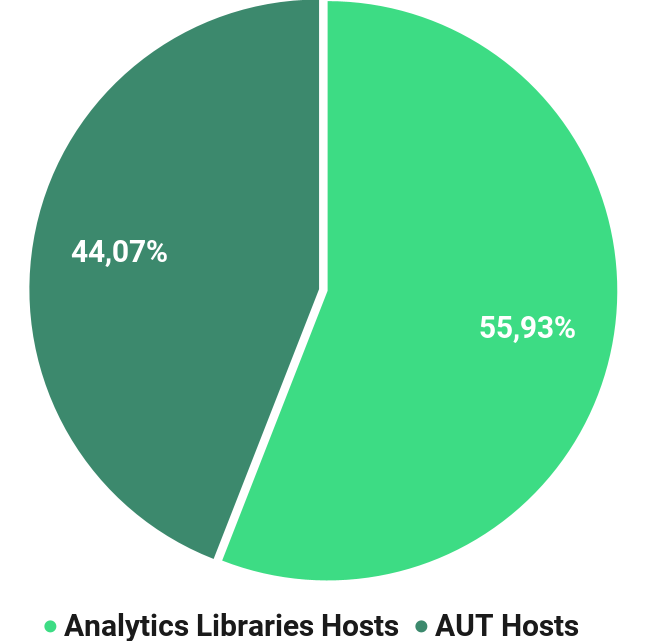}\label{fig:cake_app_hosts}}
    \subfloat[]{\includegraphics[width=0.5\linewidth]{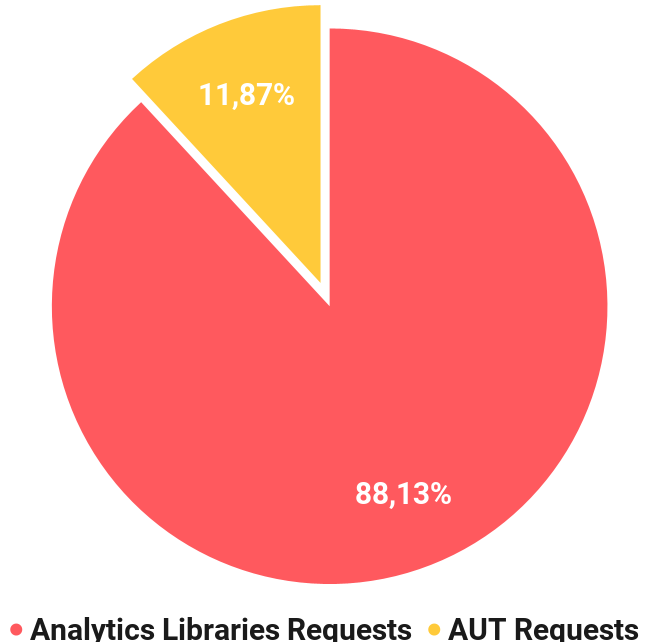}\label{fig:cake_packets}}
    
    \caption{Impact of analytics libraries on the network traffic in terms of (a) contacted hosts and (b) generated network requests.}
    \label{fig:impact_analytics_libraries_net}
\end{figure}

Figure \ref{fig:cake_app_hosts} details the impact of analytics libraries in terms of contacted hosts (a) and number of network requests (b).
In particular, among $1482$ unique network hosts, only $653$ (i.e., the $44.07\%$) are related to the normal behavior of the AUTs, while the remaining $829$ (i.e., the $55.93\%$), are connected to an analytic service (Figure \ref{fig:cake_app_hosts}).
Furthermore, it is worth noticing that over $88.13\%$ of the resulting requests (i.e., $234228$ out of $265770$) are related to analytics services (Figure \ref{fig:cake_packets}), confirming that analytics frameworks have a significant impact on the overall network traffic of apps (\textbf{RQ2}). 

\begin{figure}[t]
    \centering
    \includegraphics[width=0.7\linewidth]{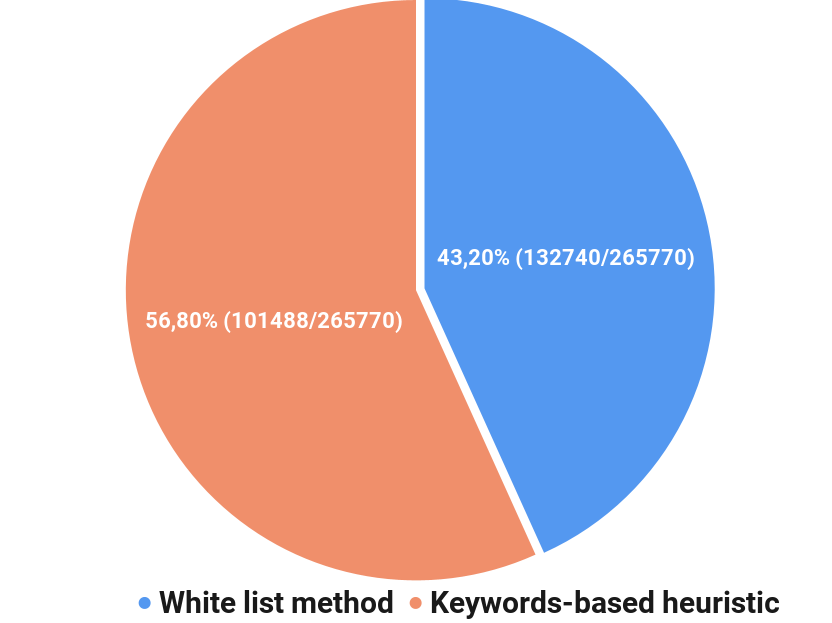}
    \caption{Impact of the two classification methods in the identification of network requests of analytics libraries.}
    \label{fig:cake_packets_heuristic}
\end{figure}

Figure \ref{fig:cake_packets_heuristic} depicts the contribution of our heuristic in the identification of network requests associated with analytics services with respect to the traditional white-list methodology. In detail, the experimental campaign allowed the identification of 132.808 new requests, representing $56.70\%$ of the total, and the mapping of 576 new hosts with the corresponding service.


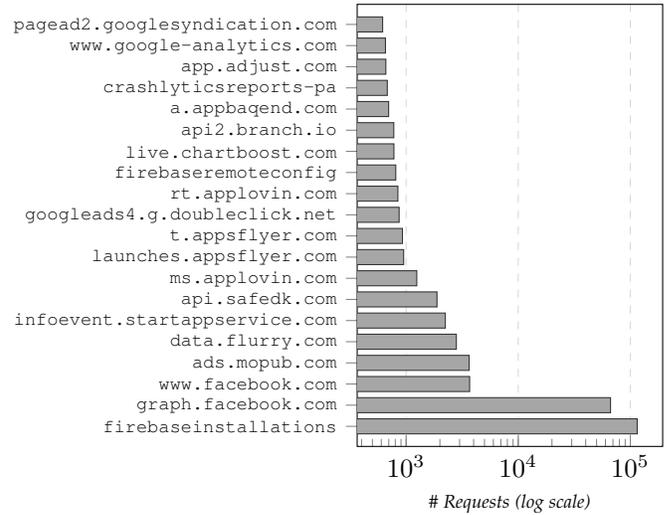
\begin{figure}[t]
    \centering
  \begin{tikzpicture}
        \begin{axis}[
        width=.31\textwidth,
        height=0.41\textwidth,
        xmajorgrids,
        grid style={dashed,gray!30},
        xbar,
        bar width=0.2cm,
        y dir=reverse,
        xmode=log,
        ytick=data,
        xmin=0,
        xlabel=\scriptsize\emph{\# Requests (log scale)},
        y tick label style={align=right,font=\scriptsize\ttfamily},
        enlarge y limits=0.05,
        tick pos=bottom,
        symbolic y coords={pagead2.googlesyndication.com, www.google-analytics.com, app.adjust.com, crashlyticsreports-pa, a.appbaqend.com, api2.branch.io, live.chartboost.com, firebaseremoteconfig, rt.applovin.com, googleads4.g.doubleclick.net, t.appsflyer.com, launches.appsflyer.com, ms.applovin.com, api.safedk.com, infoevent.startappservice.com, data.flurry.com, ads.mopub.com,www.facebook.com,graph.facebook.com,firebaseinstallations,}
        ]
        \addplot[gray!40!black,fill=gray!70!white] coordinates {
            (617,pagead2.googlesyndication.com)
            (654,www.google-analytics.com)
            (659,app.adjust.com)
            (679,crashlyticsreports-pa)
            (698,a.appbaqend.com)
            (776,api2.branch.io)
            (778,live.chartboost.com)
            (808,firebaseremoteconfig)
            (844,rt.applovin.com)
            (866,googleads4.g.doubleclick.net)
            (927,t.appsflyer.com)
            (950,launches.appsflyer.com)
            (1244,ms.applovin.com)
            (1885,api.safedk.com)
            (2237,infoevent.startappservice.com)
            (2805,data.flurry.com)
            (3648,ads.mopub.com)
            (3685,www.facebook.com)
            (66587,graph.facebook.com)
            (115217,firebaseinstallations)
        };
        \end{axis}ca
        \end{tikzpicture}
        \captionsetup{justification=centering}
        \caption{Distribution of the network requests related to analytics libraries.}\label{fig:barplot_host_packets_v2}
\end{figure}

Moreover, we identified the top 20 analytics hosts contacted during the dynamic analysis phase to further confirm the results obtained through the static analysis. Indeed, the $77\%$ of network requests (i.e., $181804$) belong to \textit{firebaseinstallations.googleapis.com} (i.e., Google) and \textit{graph.facebook.com} (i.e., Facebook) (see Figure \ref{fig:barplot_host_packets_v2}).

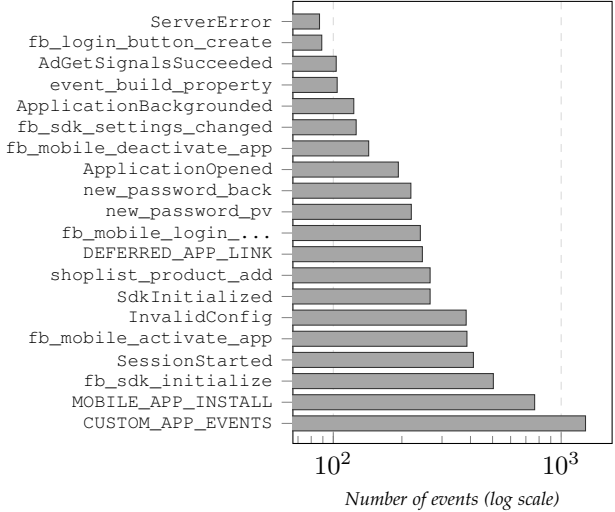
\begin{figure}[t]
    \centering
  \begin{tikzpicture}
        \begin{axis}[
        width=.32\textwidth,
        height=0.41\textwidth,
        xmajorgrids,
        grid style={dashed,gray!30},
        xbar,
        bar width=0.2cm,
        y dir=reverse,
        xmode=log,
        ytick=data,
        xmin=0,
        xlabel=\scriptsize\emph{Number of events (log scale)},
        y tick label style={align=right,font=\scriptsize\ttfamily},
        enlarge y limits=0.05,
        tick pos=bottom,
        symbolic y coords={ServerError,fb\_login\_button\_create,AdGetSignalsSucceeded,event\_build\_property,ApplicationBackgrounded,fb\_sdk\_settings\_changed,fb\_mobile\_deactivate\_app,ApplicationOpened,new\_password\_back,new\_password\_pv,fb\_mobile\_login\_...,DEFERRED\_APP\_LINK,shoplist\_product\_add,SdkInitialized,InvalidConfig,fb\_mobile\_activate\_app,SessionStarted,fb\_sdk\_initialize,MOBILE\_APP\_INSTALL,CUSTOM\_APP\_EVENTS}
        ]
        \addplot[gray!40!black,fill=gray!70!white] coordinates {
            (87,ServerError)
            (89,fb\_login\_button\_create)
            (103,AdGetSignalsSucceeded)
            (104,event\_build\_property)
            (123,ApplicationBackgrounded)
            (126,fb\_sdk\_settings\_changed)
            (143,fb\_mobile\_deactivate\_app)
            (193,ApplicationOpened)
            (219,new\_password\_back)
            (220,new\_password\_pv)
            (241,fb\_mobile\_login\_...)
            (246,DEFERRED\_APP\_LINK)
            (266,shoplist\_product\_add)
            (266,SdkInitialized)
            (383,InvalidConfig)
            (386,fb\_mobile\_activate\_app)
            (412,SessionStarted)
            (504,fb\_sdk\_initialize)
            (765,MOBILE\_APP\_INSTALL)
            (1279,CUSTOM\_APP\_EVENTS)
        };
        \end{axis}
        \end{tikzpicture}
        \captionsetup{justification=centering}
        \caption{Distribution of analytics events recorded during the dynamic analysis phase.}\label{fig:event_values}
\end{figure}

Concerning the events stored by these libraries, Figure \ref{fig:event_values} reports the set of events recorded during the dynamic analysis phase, sorted in decreasing order of frequency. The most frequent event is \textit{"CUSTOM\_APP\_EVENTS"}, belonging to the \textit{graph.facebook.com} analytics service. Through this attribute, developers can define custom events.

The evaluation of the privacy impact of analytics libraries (\textbf{RQ3}) required an in-depth review of the content of the network requests.
For each request, we extracted all the attribute keys of the event. Then, we ordered the list of attributes according to the frequency of appearance, and we classified them by evaluating their privacy impact (i.e., EI, QID, SD). The analysis of the dataset generated by the dynamic analysis phase led to the extraction of $6025$ unique attributes. 
Each attribute has been manually evaluated according to 1) its content, 2) the ability to identify the user,  and 3) its role inside the whole event.

Table \ref{tab:keyowrds_analytics_request} lists the 44 most used attributes. For each entry, we describe the attribute, an example of its usage, the privacy classification (i.e., EI, QID, SD), the number of occurrences, and the number of network hosts that received such an attribute.

The extraction and classification phase highlighted the presence of:

\begin{itemize}
    \item a set of EI that enables the unique identification of app installations (e.g., \textit{appID} and \textit{fid}), users (e.g., \textit{uid} and \textit{uuid}), or devices (e.g., \textit{device\_id} and \textit{hardware\_id});
    
    \item a set of QI that may allow for the indirect identification of the user. Examples of QI include \textit{location}, \textit{SSID} \textit{Device Name}, and \textit{Device Manufacturer};
    
    \item a set of SD related to the events generated by the interaction with the AUT that allow developers to infer the user behavior. Those attributes include \textit{name}, \textit{type}, \textit{duration}, \textit{events}, \textit{event\_type}, and \textit{event}.
\end{itemize}

Table \ref{tab:percentage_data_privacy_relevant_on_each_requests} points out the percentage of how often a request sends a data classified as privacy relevant (see Table \ref{tab:keyowrds_analytics_request}) to the most used analytics hosts.
It is worth pointing out that each request contains at least one EI, as well as that all requests expose - without any form of anonymization - at least one QID.

Overall, the privacy analysis on the information extracted by analytics libraries allowed us to empirically demonstrate their relevance for the user's privacy. 
Also, it is worth noticing that the experimental results provide no evidence that analytics services apply any anonymization or application-level encryption on the collected data.
As a consequence, there is the need for a viable and scalable anonymization strategy that grants a fine-grained user control over personal data, thereby ensuring compatibility with the existing mobile analytics solutions.

\begin{table}[t]
\tiny
    \renewcommand\arraystretch{1.35}
    \centering
    \begin{tabular}{||l|c|c|c|c||}
        \hline
        \textbf{Host} & \textbf{EI \%} & \textbf{QID \%} & \textbf{SD \%} & \textbf{\# Req}\\
        \hline 
        firebaseinstallations.googleapis.com & 99\% & 100\% & 0.1\% & 115217 \\\hline
        graph.facebook.com & 1.9\% & 100\% & 3\% & 66587\\\hline
        www.facebook.com & 95\% & 100\% & 95\% & 3685 \\\hline
        ads.mopub.com & 97\% & 100\% & 6.4\% & 3648 \\\hline
        firebaseremoteconfig.googleapis.com & 41\% & 100\% & 1.3\% & 2805 \\\hline
        rt.applovin.com & 0.04\% & 100\% & 0\% & 2237 \\\hline
        googleads4.g.doubleclick.net & 0.1\% & 100\% & 2\% & 1885 \\\hline
        t.appsflyer.com & 100\% & 100\% & 100\% & 1244\\\hline
        launches.appsflyer.com & 0.6\% & 100\% & 0.1\% & 950 \\\hline
        ms.applovin.com & 1.6\% & 100\% & 0\% & 927 \\\hline
    \end{tabular}
    \caption{Percentage of EI, QID, SD for each request in the most widespread hosts.}
    \label{tab:percentage_data_privacy_relevant_on_each_requests}
\end{table}

\newcolumntype{P}[1]{>{\arraybackslash}p{#1}}
\newcolumntype{M}[1]{>{\centering\arraybackslash}m{#1}}
\newcolumntype{?}{!{\vrule width 0.75pt}}
\begin{table*}[htb]
    \begin{adjustwidth}{-0.5cm}{}

    \centering

    {\tiny
    \begin{tabular}{M{0.7cm}M{2.2cm}M{2.6cm}M{0.3cm}M{0.5cm}M{0.6cm}?M{0.7cm}M{2.5cm}M{2.6cm}M{0.3cm}M{0.5cm}M{0.6cm}}
    
    \toprule
    \textbf{Keyword} & \textbf{Description} & \textbf{Example} & \textbf{Type} & \textbf{\# Occ} & \textbf{\# Hosts} &  \textbf{Keyword} & \textbf{Description} & \textbf{Example} & \textbf{Type} & \textbf{\# Occ} & \textbf{\# Hosts} \\
    \midrule

    \textit{appID} & Identifies a specific installation of an app, mainly used by Google libraries & ``appId": `` 1:344560015735:android:482ed113bf00cb81" & EI & 118321 & 1223 &
    \textit{uid} & User Identifier & "uid":"1609497426061\--1356139383227225046" & EI & 2064 & 10 \\ \midrule

    \textit{fid} & FirebaseID, identifies a specific installation of an app & ``fid":``cUSx\-9iNuTY\-2u8xAUm\-9tkA2" & EI & 115211 & 1045 &
    \textit{event\_type} & Specific event, generated by a user & "event\_type" : "ClickTopProductCard" & SD & 2046 & 23 \\ \midrule

    \textit{name} & Specific event, generated by a user & ``name" = ``fb\_app\_events\_enabled" & SD & 26912 & 288 &
    \textit{user} & Information about user and device & "user":\{ "deviceId" : "66dfda9e\--xxxx\--xxxx\--xxxx\--fd4909f19131", ...\} & EI & 1923 & 147 \\ \midrule

    \textit{type} & Used to identify a specific event & "type": ``perf.startupTime.v1" & SD & 11262 & 329 &
    \textit{udid} & User Identifier, used by mopub & "udid" :"mopub:ca802221\--f769\--4909\--aef7\--46c80be7353b" & EI & 1889 & 127 \\ \midrule

    \textit{model} & Specifies the device used & ``model": ``Android SDK built for x86" & QID & 11051 & 486 & 
    \textit{device\_id} & Device Unique Identifier & "deviceId":"ffffffff\--b626\--4582\--a9f2\--20d36d7a4fe6" & EI & 1887 & 125 \\ \midrule

    \textit{device} & Device name & ``device": ``generic\_x86" & QID & 7548 & 423 & 
    \textit{android\_id} & ANDROID\_ID & "android\_id": "54399037579f251d" & EI & 1662 & 69 \\ \midrule

    \textit{locale} & Device language & ``locale": ``en\_US" & QID & 6166 & 319 & 
    \textit{operator} & Mobile Telephone Operator & "operator": "T-Mobile" & QID & 1610 & 81 \\ \midrule

    \textit{manufacturer} & Device manufacturer &  ``manufacturer":``Google" & QID & 6141 & 261 & 
    \textit{userId} & Identifier associated with the device & "userid" : "02903474b\-885d424d9\-a61ea307d\-c850a" & EI & 1502 & 85 \\ \midrule

    \textit{os} & OS Version and related Information & ``os": \{``version":``10", ``buildVersion":``REL", ...\} & QID & 5230 & 379  &
    \textit{location} & Identifies the location where the event has been generated & "location": "Home Screen" & SD & 1492 & 55 \\ \midrule

    \textit{carrier} & Information about OS and Mobile TelephoneOperatore & ``carrier": \{``carrier-name" : ``Android"`, ``iso-country-code":``us",...\} & QID & 5023 & 319  &  \textit{deviceData} & Information about device hardware & "deviceData": \{"cpu\_abi" : "x86", ...\}" & EI & 1476 & 61 \\ \midrule

    \textit{current} & ID of the current event & ``current":``15" & QID & 4683 & 9 & 
    \textit{uuid} & Unique identifier generated by analytics services, used to identify a device & "uuid": "2bdc76f3-63d1-4184-93bd-060918ae6bea" & EI & 1171 & 133 \\ \midrule

    \textit{timezone} & Timezone & ``timezone" : ``Asia/Kabul" & QID & 4645 & 261 & 
    \textit{appInstanceId} & App Instance & "appInstanceId": "dVjk1CTB9kU" & QID & 1161 & 252 \\ \midrule

    \textit{app} & Identify a specific app installation. Used by Google libraries & "app": \{ "installation\-Uuid":"cad471b7\-2b064bd4b29\-e5d81201\-61f94",... \} & QID & 4427 & 237 & 
    \textit{ad\_id} & Google Advertiser Id & "ad\_id": "66dfda9e-9236-459f-9d9c-fd4909f19131" & EI & 1160 & 100 \\\midrule

    \textit{language} & Device Lamguage & "language"="en" & QID & 4152 & 367 & 
    \textit{installationUuid} & Hash relativo ad un'applicazione installata sul dispositivo & "installationUuid": "4882882\-d9829459e89d\-860c59..." & QID & 1002 & 97  \\ \midrule

    \textit{country} & Country  & "country":"IT" & QID & 4089 & 234 & 
    \textit{hardware\_id} & ANDROID\_ID & "hardware\_id":"033ae9\-5da00\-855\-66" & EI & 903 & 64 \\ \midrule

    \textit{carrierName} & OS \& Mobile Telephone Operator & "carrierName" = "Android" & QID & 3562 & 71 &
    \textit{identity} & Base 64 of uuid and gaid & "identity" : "eyJ1dWlk\-IjoiMTc1Zj\-Y2OWMwYjVhY\-jc2ZCIsImdh..." & EI & 834 & 24 \\ \midrule

    \textit{data} & Additional Information about an event & "data":\{" requests\_count ":32, "events\_count ":91, "attributes\_count":18 \} & QID & 3371 & 152 & \textit{bssid} & Mac Address of AP  & "bssid": "02:00:00:00:00:00" & EI & 738 & 15 \\ \midrule

    \textit{advertiserId} & Advertiser User Id & "advertiserId" : "8e83d747-xxxx-xxxx-xxxx-..."& EI & 3353 & 107 &  
    \textit{event} & Event Name & "event": MOBILE\_APP\_INSTALL & SD & 675 & 45 \\ \midrule

    \textit{duration} & Event duration & "duration" : 11270 & SD & 2945 & 59 & 
    \textit{deviceIP} & Device IP & "deviceIP":"xxx.0.2.xx" & EI & 671 & 6 \\ \midrule

    \textit{mid} & Mopub Generated Identifier & "mid":"8bfbff85-xxxx-xxxx-xxxx-a488bae44e1f" & EI & 2931 & 100 & 
    \textit{ssid} & SSID name & "ssid":"$<$unknown ssid$>$" & QID & 622 & 12 \\ \midrule

    \textit{events} & List of information associated to a specific events & events: \{"type": "deviceInfo", "ts" : 1607083813320, "os\_ver" : "Android OS 10 \/ API-29", ... & SD & 2656 & 296 & \textit{mac\_address} & Device Mac Address & mac\_address="yy:\-xx:\-zz:\-yy:\-xx:\-yy"& EI & 506 & 37 \\ \midrule

    \textit{network} & Network Informaion & "network": "MOBILE" & QID & 2568 & 115 &  
    \textit{deviceFingerPrintId} & Device Fingerprint & "deviceFingerPrintId": "ffffffff-bf43-71d1-ffff-ffffef05ac4a" & EI & 505 & 22  \\ \bottomrule

    \end{tabular}
    }
    \caption{Evaluation of the most privacy-relevant attributes collected by analytics libraries.}
    \label{tab:keyowrds_analytics_request}
\end{adjustwidth}
\end{table*}

\section{MobHide local anonymization}\label{sec:mobhide-hidedroid}

The need to determine the feasibility of applying user-centric anonymization techniques to the information collected by analytics libraries (\textbf{RQ4}) drove us to extend our previous work on local data anonymization on mobile devices \cite{mobhide}.
In detail, we refined the \textbf{\methodname{}} per-app anonymization methodology to cope with the state-of-the-art mobile analytics frameworks. Furthermore, we extended the \textbf{\toolname{}} prototype for the Android ecosystem to perform extensive and in-the-wild analysis on real Android apps.
The rest of this section summarizes the \methodname{} methodology, presents the extension of the \textbf{\toolname{}} prototype, and discusses the results of the experimental activity on the dataset of $4500$ Android apps.


\subsection{\methodname{}}\label{sec:sub-mobhide}

The \methodname{} methodology allows the user to choose a different privacy level for any app installed on the device. 
The idea is to dynamically analyze the app behavior at runtime and anonymize the data exported by analytics libraries.
In detail, \methodname{} leverages runtime monitoring of any app according to the following steps: i) intercept all the events generated by the analytics libraries, ii) anonymize the information therein by applying generalization and local DP techniques, and iii) send the anonymized data to the backend by mimicking the original network calls.
 
\begin{figure*}[h]
    \centering
    \includegraphics[width=0.6\textwidth]{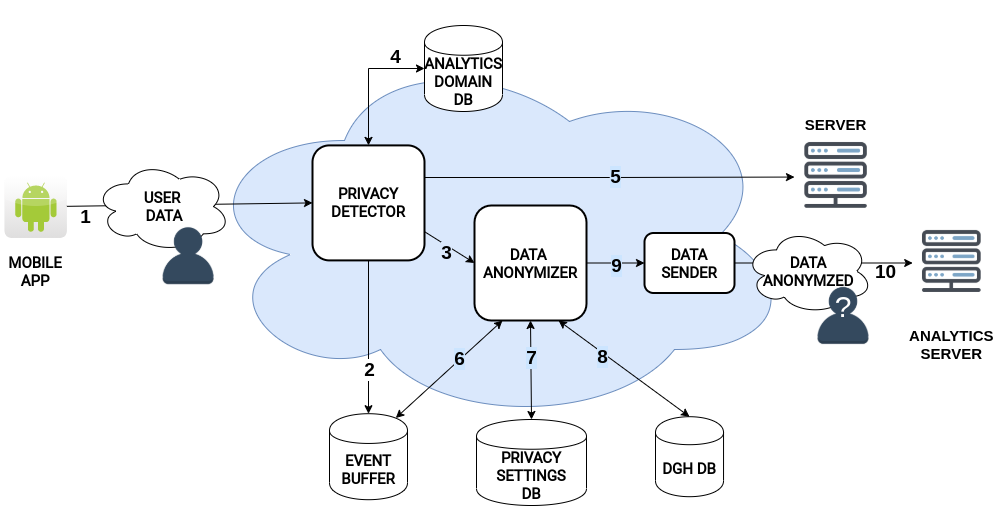}
    \captionsetup{justification=centering}
    \caption{\methodname{} Methodology.}
    \label{fig:mobhide_methodology}
\end{figure*}

However, the analysis on analytics libraries discussed in Section \ref{sec:analytics-in-the-wild} led us to revise and extend the methodology (and the prototype) originally proposed in \cite{mobhide}.

First, the previous methodology relied on a predefined list of well-known hosts of the analytics services. However, the difficulty in maintaining an updated list led to several false negatives in the preliminary experimental results.

Thus, we revised the \methodname{} methodology by: i) extending the Privacy Detector Module with a keyword-based heuristic (c.f. Section \ref{sec:analytics-in-the-wild}) and ii) including a dedicated support element, called \textit{Analytics Domain DB}, that can dynamically update the mapping among the hostnames and the corresponding analytics services.

Also, the previous proposal envisaged the storage of the intercepted network requests in a buffer, carrying out the local DP anonymization and subsequently their forwarding in blocks of $n$ requests.
Thus, the original \toolname{} prototype needed to intercept and hang a pool of open connections to analytics services to reach the desired block of requests to trigger the anonymization process.
Unfortunately, the preliminary experimental evaluation showed that such behavior is hardly achievable in a real scenario.
In detail, \toolname{} was unable to collect a sufficient number of events before the poll of connections expires due to protocol timeouts.
Furthermore, the idea of dropping the original connections and starting new ones after the anonymization process was ineffective since the analytics clients within the app tried to re-connect and re-send the same set of events, resulting in an erroneous behavior and, in some cases, to the crash of the app.

To overcome this problem, we also revised the \methodname{} anonymization pipeline, where each request is intercepted, anonymized, and forwarded sequentially. 
In detail, the new methodology exploits the local DP technique by defining a perturbation function $R$ (see Section \ref{sec:anonymization}) on the history of previous events for the same hostname rather than on a block of new intercepted events. 

Figure \ref{fig:mobhide_methodology} provides a high-level view of the new \methodname{} workflow.
The \textit{Privacy Detector} module (step 1) intercepts the network requests generated by the user apps and detects those referring to analytics services exploiting a list of well-known network hosts connected to analytics libraries and the heuristics described in Section \ref{sec:analytics-in-the-wild}.
If the network request does not belong to an analytics library, the module transparently forwards it to the destination server (step 5).
Otherwise, the \textit{Privacy Detector} stores the request in the Event Buffer module (step 2), sends the data within the request to the \textit{Data Anonymizer} (step 3), and updates the domain name in the \textit{Analytics Domain DB} (step 4).
The \textit{Data Anonymizer} is the core module of the \methodname{} methodology and it is responsible for the anonymization task. The complete procedure implemented in the \textit{Data Anonymizer} is described in Algorithm \ref{alg:pipeline-anonimization}.

The algorithm takes as input the intercepted request (i.e., $currReq$), the privacy level chosen by the user (i.e., $currPL$), the minimum number of requests (i.e., $minLen$) to carry out the DP anonymization, and the data stored in the event Buffer. 
For each request intercepted by the \textit{Privacy Detector},  the algorithm extracts the package name of the app (i.e., $appName$) and the destination server (i.e., $hostName$) (rows 1-2). Using these values, the algorithm extracts all requests between these two entities from the $event Buffer$ (row 3).

Then, the module initializes the output list of anonymized requests (i.e., $anonymizedRequests$) (row 4) and the threshold value (i.e., $Threshold_{action}$) (row 5), defined as:

\begin{equation}\label{eq:threshold-action}
    Threshold_{action} = 1 - \frac{currPL}{\# action + 1}
\end{equation}
where 
\begin{equation*}
    action \in [\texttt{inject},\texttt{remove}, \texttt{replace}].
\end{equation*}

$Threshold_{action}$ is used to determine the possible actions of the perturbation function $R$ (i.e., $inject$, $remove$ or $replace$ as defined in \cite{mobhide}).

The next step consists in the generalization process of the original request, i.e., $generalizeData$ (row 6).
Then, if the history of requests between two endpoints is above $minLen$ (row 7), the \textit{Data Anonymizer} applies the local DP anonymization and computes the three pseudo-random numbers, i.e., $Pr_{inj}, Pr_{rem}, Pr_{rep}$, used by the perturbation function $R$ to inject, remove or replace the event, respectively (rows 8-10).
If $Pr_{inj}$ is higher than the threshold (row 11), the \textit{Data Anonymizer} module picks a random generalized event from the history (row 12) and adds the new request to the $anonymizedRequests$ list (row 13) along with $currReq$.
If $Pr_{rep}$ is greater than the threshold (row 16), the module replaces the original event with one extracted from the history, i.e., $replReq$ (row 17), generalizes it (following the rules described in \cite{mobhide} and using the information stored in the \textit{DGH DB}, step 8) (row 18) and adds $replReq$ to the $anonymizedRequests$ list (row 19).
In case $Pr_{rem}$ is greater than the threshold (row 20), the \textit{Data Anonymizer} module removes the original request (row 21).

Once the request has been anonymized, the \textit{Data Anonymizer} forwards the $anonymizedRequests$  to the \textit{Data Sender}  (step 9). The \textit{Data Sender} assembles the new anonymized network request and forwards it to the analytics backend (step 10).

\begin{algorithm}[h] \caption{Data Anonymization Pipeline}\label{alg:pipeline-anonimization}
\begin{algorithmic}[1]
\footnotesize
\Require $currReq$, $currPL$, $eventBuffer$, $minLen$
\Ensure $anonymizedRequests$
\State 
$appName \gets currReq$.appName
\State  $hostName \gets currReq$.hostName
\State  $history \gets $ \parbox[t]{.6\textwidth}{$eventBuffer$.extractReq($appName,hostName$)}
\State Initialize $anonymizedRequests \gets list()$
\State Initialize $Threshold_{action} \gets 1 - (currPL/4)$

\State $currReq.att \gets$ \parbox[t]{.8\linewidth}{generalizeData($currReq.att,currPL$, $history$)}

\If{len($history$) $\geq minLen$}
    \State $Pr_{inj} \gets$ rand()
    \State $Pr_{rem} \gets$ rand()
    \State $Pr_{rep} \gets$ rand()
    \If{$Pr_{inj}$ $>$ $Threshold_{action}$}
       \State $newReq \gets$ \parbox[t]{.6\textwidth}{genNewRequest($currPL$, $history$)}
       \State $anonymizedRequests$.add($newReq$)
       \State $anonymizedRequest$.add($currReq$)
    \EndIf
    
    \If{$Pr_{rep}$ $>$ $Threshold_{action}$}
        \State $ replReq\gets$ replaceRequest($currReq$)
        \State $replReq.att \gets$ \parbox[t]{.8\textwidth}{generalizeData($replReq.att$, $currPL$, $history$)}
        \State $anonymizedRequests$.add($replReq$)

    \ElsIf{$Pr_{rem}$ $>$ $Threshold_{action}$}
        \State deleteEvent($currReq$)
    
    \Else
        \State $anonymizedRequest$.add($currReq$)        
        
    \EndIf
\Else
   \State $anonymizedRequest$.add($currReq$)
\EndIf
\State \Return $anonymizedRequests$
\end{algorithmic}
\end{algorithm}

\subsection{\toolname{}}

\textbf{\toolname{}} implements the \methodname{} methodology for the Android ecosystem as a user app compatible with Android 6.0 and above. The application, after an initial configuration (\textbf{Initial Setup}), enables users to select a privacy level for each of the installed apps (\textbf{Per-App Privacy Configuration}) and, thanks to an embedded network proxy, allows the traffic collection and anonymization phase (\textbf{Runtime Anonymization}). 
\toolname{} is publicly available on GitHub \cite{HideDroidGithub}.

\subsubsection{Initial Setup}\label{sec:hidedroid_setup}
\toolname{} requires an initial configuration to successfully intercept the network traffic generated by the apps. 
At the first execution, \toolname{} requires the permission to access the external storage (i.e., \texttt{WRITE\_EXTERNAL\_STORAGE}) to store the intercepted network traffic.

\begin{figure}[t]
\centering
    \subfloat[]{\includegraphics[width=0.3 \linewidth]{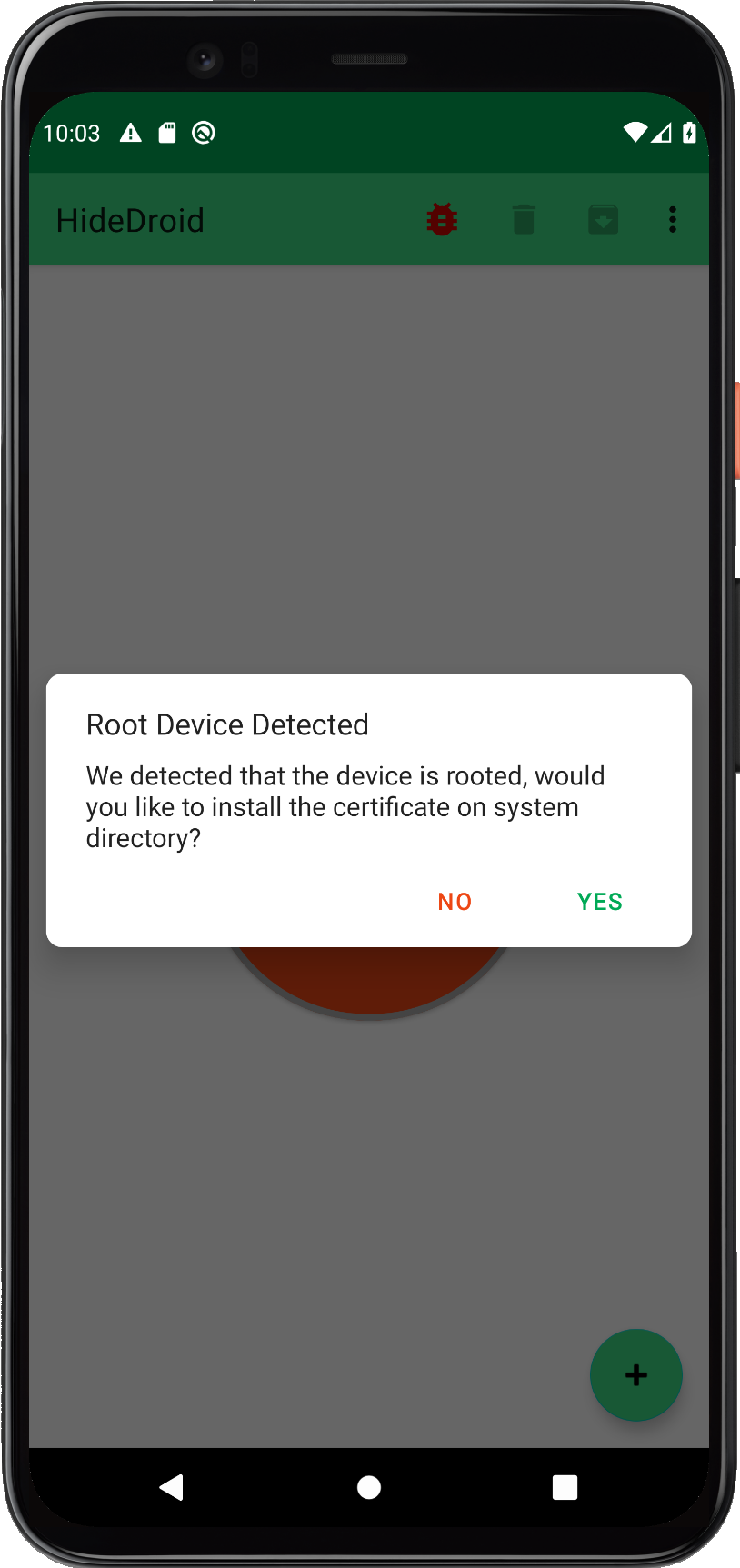}\label{fig:certificate_installation_hide_droid_a}}
    \hspace{3mm}
    \subfloat[]{\includegraphics[width=0.3 \linewidth]{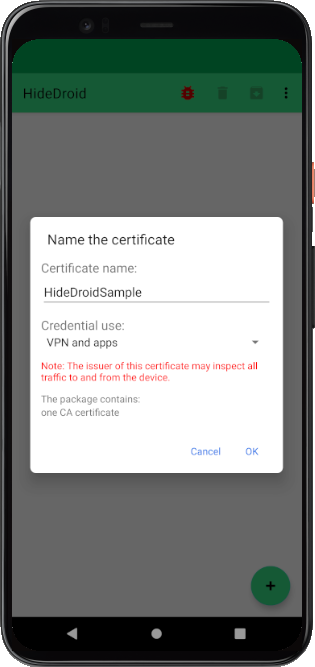}\label{fig:certificate_installation_hide_droid_b}}
    \hspace{3mm}
    \subfloat[]{\includegraphics[width=0.3 \linewidth]{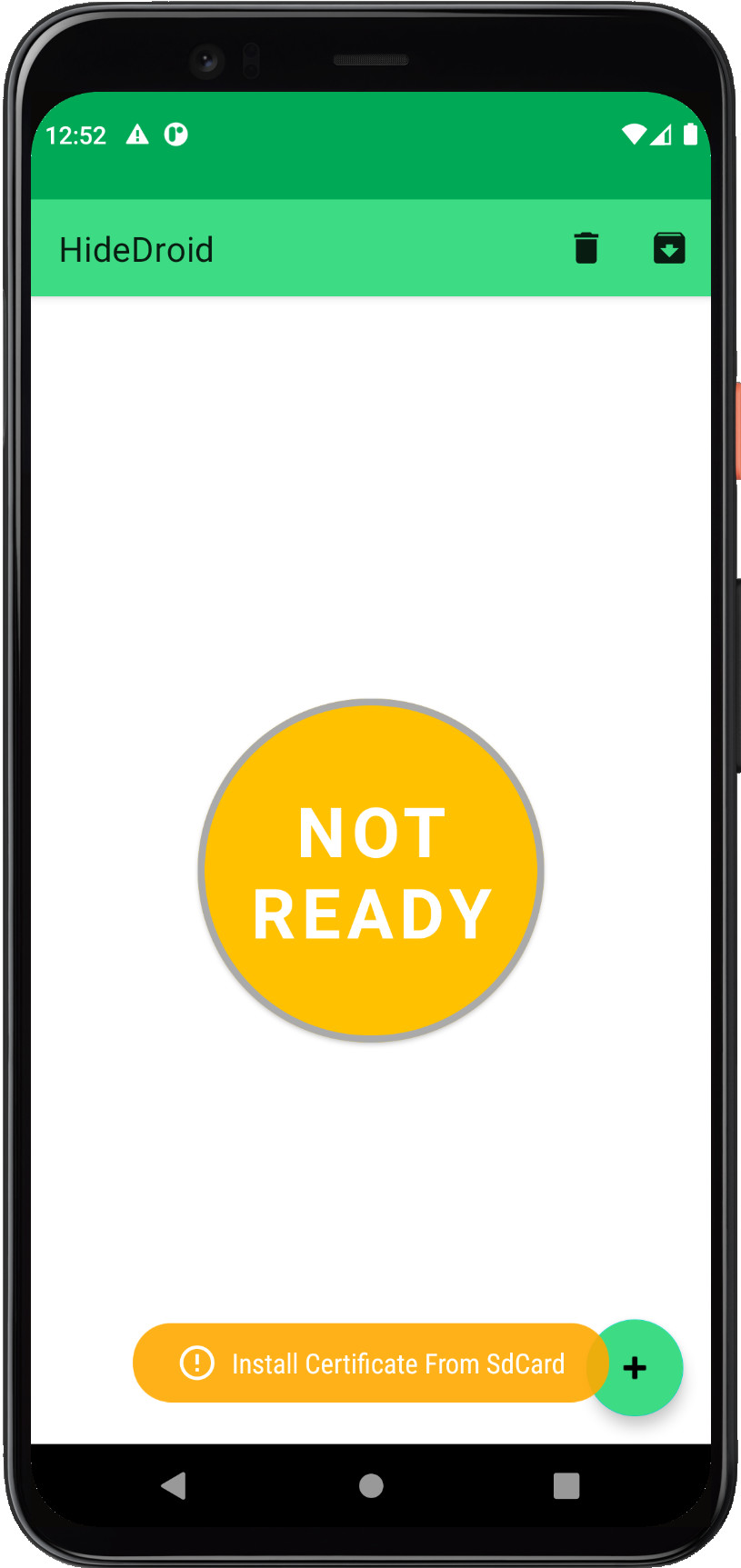}\label{fig:certificate_installation_hide_droid_c}}
    \caption{\toolname{} prompts. \textit{(a)} Root detection. \textit{(b)} In-app certificate installation (Android 10 or lower). \textit{(c)} Manual installation of the CA certificate (Android 11+).}
    \label{fig:certificate_installation_hide_droid}
\end{figure}

Then, the app generates and installs a self-signed certificate for the network traffic collection.
During such a process, \toolname{} checks if the device has root permissions. 
If this is the case, the app requests the permission to install the certificate within the system directory, which is inaccessible with the default user permissions (Figure \ref{fig:certificate_installation_hide_droid_a}) \cite{cacertwiki}. The certificate installation in the system directory allows \toolname{} to bypass an extra configuration step in the next phase (i.e., the \textit{repackaging phase}).
In case the device has default permissions only, \toolname{} executes two different actions based on the Android version installed on the device. If the Android version is lower than Android 11, \toolname{} prompts the user to install the proxy certificate in the user directory (Figure \ref{fig:certificate_installation_hide_droid_b}).
Otherwise, the app asks the user to install the certificate (Figure \ref{fig:certificate_installation_hide_droid_c}) manually. Such action is needed because Android 11 and above have tightened the restrictions on CA certificates, denying any app, debugging tool, or automated action to prompt the installation of a CA certificate \cite{Android_11_CA}.

\begin{figure}[t]
\centering
    \subfloat[]{\includegraphics[width=0.3 \linewidth]{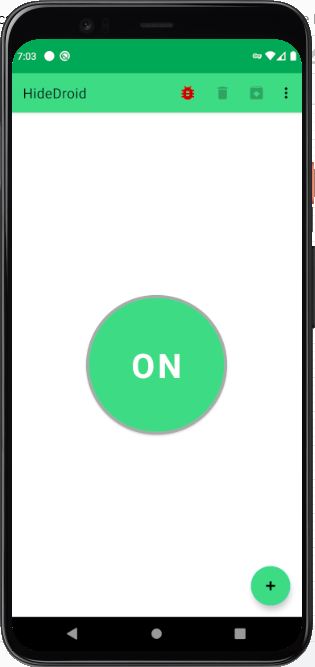}\label{fig:on_state_hide_droid}}
    \hspace{3mm}
    \subfloat[]{\includegraphics[width=0.3 \linewidth]{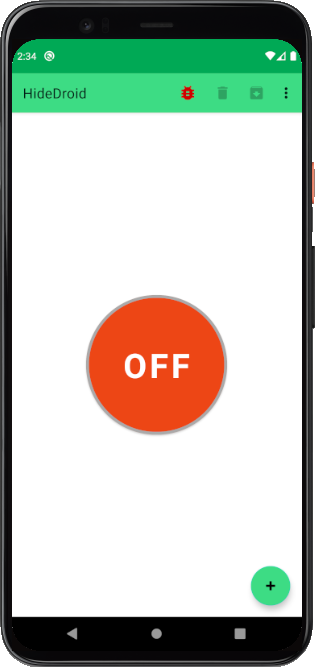}\label{fig:off_state_hide_droid}}
    \hspace{3mm}
    \subfloat[]{\includegraphics[width=0.3 \linewidth]{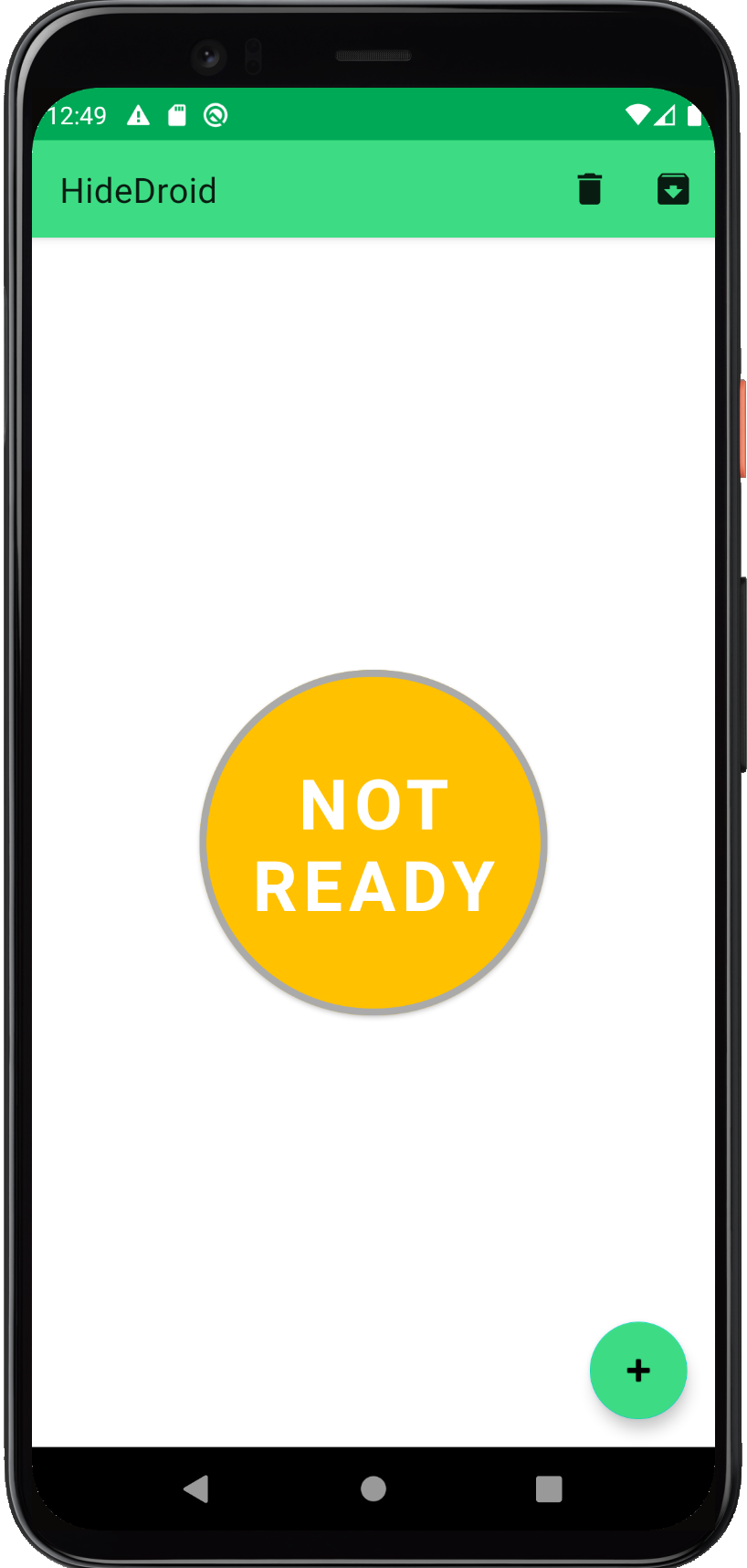}\label{fig:not_ready_state_hide_droid}}
    \caption{\toolname{} home screen. \textit{(a)} Incognito Mode on. \textit{(b)} Incognito Mode off. \textit{(c)} Not Ready.}
    
    \label{fig:start_ui_hidedroid}
\end{figure}

\begin{figure}[t]
\centering
    \subfloat[]{\includegraphics[width=0.35 \linewidth]{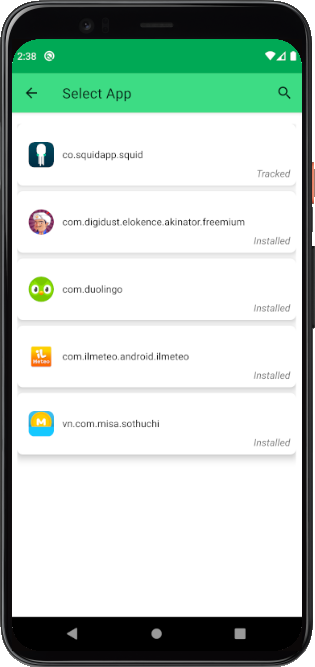}\label{fig:select_app_hide_droid}}\hspace{10mm}
    \subfloat[]{\includegraphics[width=0.35 \linewidth]{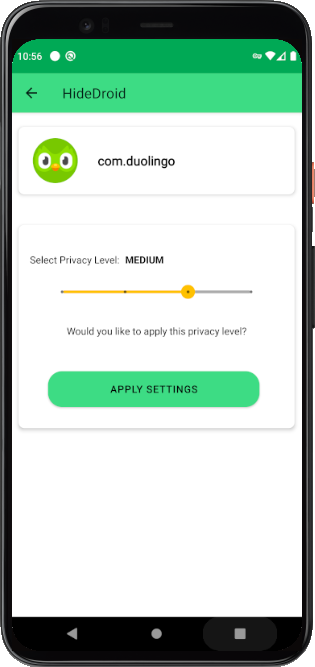}\label{fig:select_privacy_level_hidedroid}}
    
    \caption{Per-App Privacy Configuration.}
    \label{fig:configuration_app_hide_droid}
\end{figure}

\subsubsection{Per-App Privacy Configuration}
\label{sec:repackaging}

Once the setup phase has completed, \toolname{} displays the home page screen 
(Figure \ref{fig:on_state_hide_droid} and Figure \ref{fig:off_state_hide_droid}). Here,  the user can activate the anonymization mode (on/off button) and select the apps to shield (plus button). 
If this is not the case, \toolname{} displays a Not Ready warning (Figure \ref{fig:not_ready_state_hide_droid}). This screen is prompted until the certificate is successfully installed on the device.

The app selection process shows the list of all the installed apps (Figure \ref{fig:select_app_hide_droid}). For each app, the user can select the desired privacy level through a slider widget (Figure \ref{fig:select_privacy_level_hidedroid}).

\noindent
If the \toolname{} certificate is not installed in the system certificate store and the device has Android OS version $\geq 7.0$,
the tool requires the \textbf{repackaging} of each app whose selected privacy level is above \texttt{NONE}.

Such an additional step is mandatory to overcome the network security restriction imposed by the newer Android versions \cite{AndroidNougatNetwork,AndoridSecurityCOnfig} that do not recognize user certificates as trusted by default.

In detail, \toolname{} automatically unpacks each selected app, overwrites the \texttt{net\-work\_secu\-rity\_con\-fig.xml} file (Listing  \ref{lst:network-security-config-code}) to include a new trust-anchor for user-defined certificates, rebuilds the apk files, and installs the modified versions. 

It is worth noticing that - thanks to the repackaging phase - the user is not forced to have (and grant) root permissions to \toolname{}, thereby ensuring a wider compatibility w.r.t. the state-of-the-art solutions. Also, the repackaging is only required for devices with OS $\geq 7.0$ to support the newest Android OS versions without breaking their standard security model.
Finally, the modification of the app neither alters the compiled code nor other resources of the apps.

\begin{lstlisting}[caption={Example of network configuration.},
label={lst:network-security-config-code}, language=XML,float=t]
  <network-security-config>  
      <base-config>  
          <trust-anchors>  
              <!-- Trust preinstalled CAs -->  
              <certificates src="system" />  
              <!-- Additionally trust user added CAs -->  
              <certificates src="user" />  
          </trust-anchors>  
      </base-config>  
  </network-security-config>
\end{lstlisting}

\subsubsection{Runtime  Anonymization}\label{sec:data_anonymization_hidedroid}

The \textit{Data Anonymization} is the core phase of \toolname{} workflow. In detail, \toolname{} implements the \methodname{} anonymization pipeline to anonymize the traffic generated by analytics libraries of all the apps configured in the previous phase with a privacy level \texttt{LOW}, \texttt{MEDIUM}, or \texttt{HIGH}. The core of the pipeline is implemented as two background services, i.e., the  \textbf{Privacy Interceptor} and the \textbf{Data Anonymizer}.

The \textbf{Privacy Interceptor} component has a twofold objective. The first is to collect the network traffic of all apps with a selected privacy level above \texttt{NONE}. The second is to filter network requests that do not belong to any analytics services.

To intercept network requests generated by apps, the component relies on and exploits the Android VPN API \cite{vpn-service}. These APIs allow building a transparent VPN that acts like a proxy server between the client (i.e., the app) and the server (i.e., the analytics server). 
This solution enables \toolname{} to intercept the network traffic generated both from Java/Kotlin and native code, and it is able to inspect both HTTP and HTTPS traffic by exploiting SSL deep-inspection techniques.

Then, the Privacy Interceptor differentiates the traffic generated by apps according to their package names.
For Android versions lower than 10, the module leverages the \texttt{/proc/net} file to obtain the UID of the app that generated a specific request. For Android versions above 10, the Privacy Interceptor uses the \texttt{getConnectionOwnerUid}\footnote{\url{https://github.com/Mobile-IoT-Security-Lab/HideDroid/blob/main/netbare-core/src/main/java/com/github/megatronking/netbare/net/UidDumper.java}} method \cite{Android_10_privacy}. 
Finally, the module maps the UID of the process with the package name of the app using the \texttt{getPackagesForUid}\footnote{\url{https://developer.android.com/reference/android/content/pm/PackageManager\#getPackagesForUid(int)}} API.

After the collection phase, the Privacy Interceptor filters all the collected network traffic to identify the one belonging to analytics services.
At first, the module queries the \textit{Analytics Domain DB} to spot well-known domain names belonging to analytics frameworks.
If no match is found, then the service enforces the heuristic described in Section \ref{sec:analytics-in-the-wild} to evaluate the request. If the domain is recognized as belonging to an analytics service, the module blocks the request and stores it for the anonymization phase. On the contrary, the network request is transparently forwarded to the destination.\\
The \textbf{Data Anonymizer} is responsible for applying the anonymization strategies on all the stored network requests.
As a preliminary step, the service decodes each network request to preserve the original structure after the anonymization. For each request, the service extracts information regarding the headers and the body of the request, as shown in Table \ref{tab:parsed_request}. The current version of \toolname{} supports the \texttt{Content-Types} and \texttt{Content-Encodings} listed in Table \ref{tab:contenttype}. 
After the parsing phase, the Data Anonymizer can employ the anonymization process using the \textit{Generalization} and \textit{Differential Privacy} techniques following the Anonymization Pipeline of Algorithm 1. 
Finally, the anonymized request will be encoded in the original form and forwarded to the original target server by the \textit{Data Sender} module.

\begin{table}[t]
    \renewcommand\arraystretch{1.5}
    \scriptsize
    \centering
    \begin{tabular}{||c|c||}
         \hline
        \textbf{Content-Type} & \texttt{application/x\--www\--form\--url\-encoded} \\
        
        \hline
        
       \textbf{Content-Type} & \texttt{application/json} \\
        
        \hline
        
       \textbf{ Content-Type} & \texttt{multipart/form-data} \\
        
        \hline

        \textbf{Content-Encoding} & \texttt{gzip} \\
        
        \hline

        \textbf{Content-Encoding} & \texttt{deflate} \\
        
        \hline

    \end{tabular}
    \caption{Content Type and Content Encoding supported by \toolname{}.}
    \label{tab:contenttype}
\end{table}

\begin{table}[t]
    \renewcommand\arraystretch{1.5}
    \scriptsize
    \centering
    \begin{tabular}{||c|P{5.6cm}||}
        \hline
        \textbf{Headers} & Intercepted request header\\ \hline
        \textbf{Content-Type} & Intercepted request Content-Type\\ \hline
        \textbf{Content-Encoding} & Intercepted request Content-Encoding\\ \hline
        \textbf{URL} & Destination address (host and path) and request type (POST, GET, PUT, etc.)\\ \hline
        \textbf{Body} & Intercepted request body\\ \hline
        \textbf{App} & App Name\\ \hline
    \end{tabular}
    \caption{Parsing table of analytics network requests.}
    \label{tab:parsed_request}
\end{table}

\subsection{Testing \toolname{} In The Wild}\label{sec:testing_hidedroid_in_the_wild}
We conducted another experimental campaign on the same dataset of $4500$ Android apps used for the evaluation of analytics libraries in the wild (cf. Section \ref{sec:analytics-in-the-wild}), in order to evaluate the effectiveness and efficacy of \toolname{}. 
The experiments relied on an emulator equipped with Android 10 without root permissions. By using such an environment, we were able to test all steps performed by \toolname{}, including the repackaging phase, which is not mandatory in case of root permissions or Android OSes below 7.0 (see Section \ref{sec:hidedroid_setup}).
We tested each app for $10$ minutes, including the time to perform the configuration tasks (repackaging, privacy-level selection) before executing the app.

\noindent
\textbf{Runtime Performance.}
The experimental evaluation of $4500$ apps lasted one month. 
\toolname{}  was able to successfully process and anonymize data belonging to $3992$ apps (i.e., $88.7\%$). 
The remaining $508$ apps (i.e., $11.3\%$) were not tracked by \toolname{} due to the failure of the repackaging phase. The root-cause analysis of the failure allowed us to identify that the failure was triggered during the re-installation of the modified app. In detail, the error is generated by the \textit{VerifyAdvancedProtectionInstallTask} method of the \texttt{com.google.android.finsky} process. Such control verifies the app signature by comparing it with the original one. If the two signatures do not match, the process blocks the installation. As additional proof, we also tested the AUTs that failed in an Android emulator with root permissions, confirming their actual functioning.

The dynamic testing phase allowed measuring the impact of the delays introduced by the anonymization pipeline on the AUTs.
To do so, we measured the delay introduced by the interception, anonymization, and forward of each analytics event. \toolname{} was able to process, on average, an event in 52.84 ms with a standard deviation of 122.18 ms, thus, confirming a negligible impact on the AUTs.

Furthermore, we also evaluated the compatibility of \toolname{} for the leading analytics services during the experimental phase. In detail, we tracked the acceptance rate of the anonymized requests by the analytics back-end services.   
Table \ref{tab:perc_accepted_requests} shows the percentage of anonymized events accepted by the ten most used analytics services.
On average, the acceptance rate of most of the analytics services like Google ADS and Facebook is above $93.69\%$. The only notable exception is the Firebase Analytics services that systematically deny almost all the anonymized events sent by \toolname{}.
Such a limitation is due to the usage of a proprietary format called \textit{protobuf} \cite{protobuf} to serialize the network requests delivered to the back-end services. 
Unfortunately, without an a-priory knowledge of the structure of the protobuf request, it is not possible to successfully parse the data \cite{protobuf_guide}. Thus, \toolname{} is not able to correctly process the requests, causing a significant drop in the acceptance rate.
Still, it is worth noticing that the dynamic testing phase confirms that the failure in the acceptance of the anonymized events does not interfere with the normal execution of any of the AUTs.

\begin{table}[t]
    \scriptsize
        \renewcommand\arraystretch{1.35}
        \centering
        \begin{tabular}{||c|c||}
            \hline
            \textbf{Service} & \multicolumn{1}{|P{3cm}||}{\textbf{Acceptance Rate}}\\
            \hline
                Firebase Analytics & $0.19\% (317/164272)$\\ \hline
                Facebook Audience & $93.69\% (81394/86872)$\\ \hline
                Google DoubleClick & $99.46\% (85516/85982)$\\ \hline
                Google AdMob & $99.96\% (47713/47733)$\\ \hline
                Google Tag Manager & $99.88\% (6840/6848)$\\ \hline
                Google Ads & $93.94\% (5285/5626)$\\ \hline
                AppLovin & $95.69\% (4949/5172)$\\ \hline
                Twitter MoPub & $55.25\% (2227/4031)$\\ \hline
                Google CrashLytics & $98.81\% (2158/2184)$\\ \hline
                Google Analytics & $80.77\% (1684/2085)$\\\hline 
                Others & $65.81\% (28933/43961)$\\\hline
                \textbf{TOT} & $58.72\% (267016/454766)$\\ \hline
        \end{tabular}
        \caption{Acceptance rate of the anonymized events of the top 10 analytics backend services.}
        \label{tab:perc_accepted_requests}
\end{table}

\noindent
\textbf{Evaluation of the Anonymization Process.}
During the experimental phase, \toolname{} was able to anonymize more than $200k$ requests. Listing \ref{lst:example_of_request} depicts an example of a request generated by an analytics library. 
The request contains several information regarding the network connection, device, and location. This information can be divided into QID (i.e., \textit{network mode}, \textit{operator}, \textit{country}, \textit{language}, \textit{device}, \textit{model}, \textit{os}, \textit{local IP}, \textit{bssid}, and \textit{ssid}) or EI (i.e., \textit{mac address}, \textit{hardware\_id}, and \textit{device\_id}).
Moreover, the request contains also information about the event generated by the user (i.e., \texttt{AddToCart}) and additional details about it (i.e., \textit{contents}, \textit{id\_content}, \textit{price}, and \textit{content\_type}).

\begin{lstlisting}[caption={Example of an analytics request intercepted by \toolname{}.}, label={lst:example_of_request},float=t]
    {
        "hardware_id":"033ae95da0085566", 
        "brand":"Google",
        "device_id":"ffffffff-b626-4582-a9f2-20d36d7a4fe6",
        "model": "Android SDK built for x86",
        "screen_dpi": 560,
        "screen_height": 2701,
        "network": "MOBILE",
        "operator": "T-Mobile",
        "screen_width": 1440,
        "os": "Android",
        "country": "US",
        "language": "en",
        "local_ip": "10.0.2.15",
        "bssid": "02:00:00:00:00:00",
        "ssid": "You are WiFizoned",
        "mac_address": "00:10:FA:6E:38:4A",
        "latest_install_time": 1609930857411,
        "latest_update_time": 1609930857411,
        "first_install_time": 1609930857411,
        "google_advertising_id": "8e83d747-13ec-491f-89bb-761e9d0cef11",
        "sdk": "android3.0.4", 
        "branch_key": "key_live_pfv4qQtKHbRzXlt5hHufpbmgEBlgiG57",
        "event_type": "AddToCart",
        "data": {
            "contents":"ddr4 memory",
            "id_content":"34",
            "content_type":"pc hardware",
            "price":"300",
        }
    }
\end{lstlisting}

Listing \ref{lst:example_of_anonmyzed_request} represents the same network request anonymized by \toolname{} and exploiting the anonymization pipeline described in Algorithm \ref{alg:pipeline-anonimization}.
In detail, the tool relied on the DGH rules to anonymize the information regarding, e.g., brand, device model, network operator, and OS version.
All the other information about the user and the device (e.g., \texttt{hardware\_id},  \texttt{device\_id}, \texttt{local IP}, \texttt{bssid}, and \texttt{ssid}) are generalized using the generalization procedure described in \cite{mobhide}.
Finally, \toolname{} replaced the recorded event (i.e., \texttt{AddToCart}), with another one taken from the pool of events (i.e., \texttt{OpenApp}).

\begin{lstlisting}[caption={Example of an analytics request anonymized by \toolname{}.}, label={lst:example_of_anonmyzed_request},float=t]
    {
        "hardware_id": "033ae9**********", 
        "brand": "Smartphone",
        "device_id":"fffffffff-b62************************",
        "model": "Android",
        "screen_dpi": 500,
        "screen_height": 2700,
        "network": "Mobile Operator",
        "operator": "Anonymous Operator",
        "screen_width": 1400,
        "os": "Smartphone OS",
        "country": "America",
        "language": "en",
        "local_ip": "10.0.0.0",
        "bssid": "02:00:***********",
        "ssid": "You ar***********",
        "mac_address": "00:10:***********",
        "latest_install_time": 1609900000000,
        "latest_update_time": 1609900000000,
        "first_install_time": 1609900000000,
        "google_advertising_id": "8e83d747-13e************************",
        "sdk": "sdk version", 
        "branch_key": "key_live_pfv4q***************************",
        "event_type": "OpenApp"
    }
\end{lstlisting}

To evaluate the anonymization using local DP techniques, we further inspected the $150$ apps that generated most of the analytics events during the dynamic analysis phase. 
In detail, we replicated the dynamic analysis by stimulating each AUT and recording the anonymized network requests produced by \toolname{} in $10$ minutes for each of the available privacy levels.

Table \ref{tab:overview_results} reports the results of the analysis on the set of $150$ apps.  
In detail, we reported, for each privacy level, the $Threshold_{action}$ (i.e., TH), the mean number of injected, removed and replaced events (i.e., $\overline{\# Inj_{Ev}}$, $\overline{\# Rem_{Ev}}$ and $\overline{\# Rep_{Ev}}$, respectively), and the  mean number of total events  (i.e., $\overline{\# Tot_{Ev}}$). 

Finally, we computed the KL\_Divergence metric \cite{kullback1997information} ($D_{KL}$) to evaluate the anonymization process in terms of privacy and utility. This metric allows measuring the \emph{distance} between two distributions of events. A high value of $D_{KL}$ suggests that the two distributions are very different, i.e., the anonymization process overturns the original data at the expense of its utility. On the other hand, a value equals to 0 indicates that the two distributions are identical, i.e., the anonymization process does not modify the collected data, hence preserving the maximum level of utility.

The last column in Table \ref{tab:overview_results} reports the mean KL\_Divergence (i.e., $\overline{D_{KL}}$) between the original distribution of analytics events and each anonymized distribution obtained with the \texttt{LOW}, \texttt{MEDIUM}, and \texttt{HIGH} levels, respectively.

It is worth pointing out that the distance between the original distribution and the anonymized ones is always greater than 0, meaning that the local DP anonymization has successfully increased the privacy of the distribution of events.
Also, $D_{KL}$ value is always close to 0, which means that the anonymization process did not overturn the collected data, thereby preserving a reasonable level of utility.
Furthermore, the higher is the privacy level, the greater is the $D_{KL}$ value, thereby demonstrating that the utility of the exported data actually lowers when the privacy level rises.

\begin{table}[t]
    \renewcommand\arraystretch{1.5}
    \scriptsize
    \centering
    \begin{tabular}{||M{0.8cm}M{0.5cm}M{0.9cm}M{0.9cm}M{0.9cm}M{0.9cm}M{0.5cm}||}
    \hline

    \pmb{Privacy} & \pmb{TH} & \textbf{$\overline{\# \pmb{Inj_{Ev}}}$} & \textbf{$\overline{\# \pmb{Rem_{Ev}}}$} & \textbf{$\overline{\# \pmb{Rep_{Ev}}}$} & \textbf{$\overline{\# \pmb{Tot_{Ev}}}$} & \textbf{$\overline{\pmb{D_{KL}}}$} \\
    \hline
    \pmb{LOW} & $0.75$ & 8.53 & 6.44 & 8.68 & 37.18 & 0.05 \\
    \pmb{MEDIUM} & $0.5$ & 17.64 & 8.98 & 17.29 & 43.76 & 0.11 \\
    \pmb{HIGH} & $0.25$ & 26.74 & 6.09 & 26.07 & 55.74 & 0.19  \\
    \hline

    \end{tabular}
    \caption{Experimental results of LDP techniques used by \toolname{} on the $150$ apps dataset.}
    \label{tab:overview_results}
\end{table}

\section{Related Work}\label{sec:related-work}
The wide adoption of third-party analytics libraries in mobile apps has recently attracted the attention of the security research community.
The work of Chen et al. \cite{information_leakage_through_mobile_analytics_services} is one of the first studies focused on the privacy issues related to mobile analytics libraries.
In detail, the authors demonstrated how an external adversary could extract sensitive information regarding the user and the app by exploiting two mobile analytics services: Google Mobile App Analytics and Flurry.
Moreover, Vallina et al. \cite{vallina2016tracking} identified and mapped the network domains associated with mobile ads and user tracking libraries through an extensive study on popular Android apps.
The authors in \cite{zhang2020does} highlighted the privacy problem related to a misconfiguration of analytic services.
In detail, they proposed PAMDroid, a semi-automated approach to investigate whether mobile app analytic services are actually anonymous and how Attributes Setting Methods (ASMs) can be misconfigured by app developers. 
These ASMs can be misconfigured by developers so that individual user behavior profiles can be disclosed, which might impose greater privacy risks to users.
All the above-mentioned works focused on the privacy implications on the usage of analytics libraries and determined that analytics services do not apply any anonymization methodology, thereby highlighting how misconfigurations in those services by the app developers may lead to severe privacy breaches. Their work acted as a motivation for our empirical study to investigate and classify the data collected by analytics service and pushed the design of \methodname{} and \toolname{}. Also, the core of our work is to propose a sound methodology to enhance the privacy of the collected data. However, the identification of application-level or service-level privacy misconfigurations is out of the scope of this work and can be demanded to further extensions.
Most of the research activity focuses on proposing novel approaches to enhance user privacy. 
Beresford et al. \cite{alastair2011Mockdroid} proposed a modified version of the Android OS called MockDroid, which allows to "mock" the access of mobile apps to system resources. 
MockDroid allows users to revoke access to specific resources at run-time, encouraging the same users to take into consideration a trade-off between functionality and personal information disclosure. 
Zhang et al. \cite{zhangprivaid} proposed PrivAid, a methodology to apply differential privacy anonymization to the user events collected by mobile apps. The tool replaced the original analytics API with a custom implementation that collects the generated event and applies DP techniques. 
The anonymization strategy is configured directly by the app developer, which is able to reconstruct at least a good approximation of the distribution of the original events.
The authors in \cite{tracking_app_analysis} proposed an Android app called Lumen Privacy Monitor that analyzes network traffic on mobile devices. This app aims to alert the user if an app collects and sends personally identifiable information (e.g., IMEI, MAC, Phone Number). The application allows the user to block requests to a specific endpoint. To do that, Lumen Privacy Monitor asks for all the Android permissions in order to collect the user data and perform the lookup in the network requests. 
Zhang et al. \cite{zhang2019introducing} and Latif et al. \cite{latif2020introducing} evaluated the feasibility of the Differential Privacy (DP) approach in the anonymization process of dynamically-created content that is retrieved from a content server and is displayed to the app user.
They described how DP could be introduced in screen event frequency analysis for mobile apps, and demonstrated an instance of this approach for Android apps and the Google Analytics framework. 
Then, they developed an automated solution for analysis, code rewriting, and run-time processing in order to modify the original distribution of screen events preserving, however, the accuracy of the data.
Unfortunately, the above solutions do not provide proper data anonymization, thereby proposing either block-or-allow strategies or approaches that enable the reconstruction of the original data by a third-party (e.g., the app developer).
Also, most of them require invasive modifications of the apps or the OS (e.g., custom OS and root permissions), and can very hardly be adopted in the wild. 
To the best of our knowledge, this work is the first proposal that analyzes the usage of analytics libraries in the wild evaluating the real user privacy threats.
Moreover, in this work, we also extended our user-centric methodology (proposed in \cite{mobhide}) and described our prototype for Android devices.

\section{Conclusion}\label{sec:conclusion}

\noindent

In this paper, we have analyzed the widespread of analytics libraries and their impact on the privacy of the user and the device by conducting a systematic and automated analysis on the top $4500$ Android applications extracted by the Google Play Store.

The obtained results drove us to propose i) an extension of our per-app anonymization methodology - \methodname{} - and ii) a prototype implementation for the Android ecosystem - HideDroid -  to cope with state-of-the-art mobile analytics frameworks.

The results obtained by our experiments demonstrated that a user-centric solution for anonymizing data collected by analytics libraries is applicable in the wild with a negligible impact on the user and the device.

Still, we advocate that the current methodology can be extended by adopting more sophisticated techniques to adapt to the nature of the identified data.
In this respect, we plan to investigate new anonymization strategies such as CAHD \cite{ghinita2008anonymization}, l-diversity \cite{machanavajjhala2007diversity} or t-closeness \cite{li2007t}.

Also, we were able to identify several limitations in the adoption of the Android VPN APIs that are, thus, inherited by \toolname{}.
In particular, if an analytic library enforces SSL Pinning techniques to protect its network traffic, \toolname{} is not able to intercept the network requests because the Android app raises an exception due to the invalid certificate. Despite the existence of SSL bypass techniques such as the use of Frida \cite{FridaSSL}, or Xposed \cite{XposedSSL}, they either require root permissions or per-app instrumentation, which may lead to the crash of the AUT. 
Moreover, if the app developer applies an additional encryption layer on the network traffic, \toolname{} will not be able to decrypt the data programmatically. We mitigated such issues in \toolname{} by considering encrypted data as generic strings even though the corresponding anonymization process (e.g., the data generalization) would break the decryption process at the backend side. The rationale of such a choice is to prioritize the privacy of the collected information with respect to the utility. 
To overcome the limitation of such technologies, we plan to investigate the use of virtual environment technologies, such as VirtualApp \cite{virtualapp} and DroidPlugin \cite{droidplugin}, that enable the dynamic hooking of all the events generated by analytic libraries without the need to modify and repack the application.
Moreover, by using a virtualization-based approach, we could investigate the extension of \methodname{} and \toolname{} to all the data collected by apps that could affect the privacy of the user.


%

\ifCLASSOPTIONcompsoc
  \section*{Acknowledgments}
\else
  \section*{Acknowledgment}
\fi

This work was partially funded by UNIGE Starting Grant project ``User-defined Data Privacy in Android" (2018).

\ifCLASSOPTIONcaptionsoff
  \newpage
\fi



%
\bibliographystyle{IEEEtran}
\bibliography{bibliography}

%
\vspace{-1cm}
\begin{IEEEbiography}[{\includegraphics[width=1in,height=1.25in,clip,keepaspectratio]{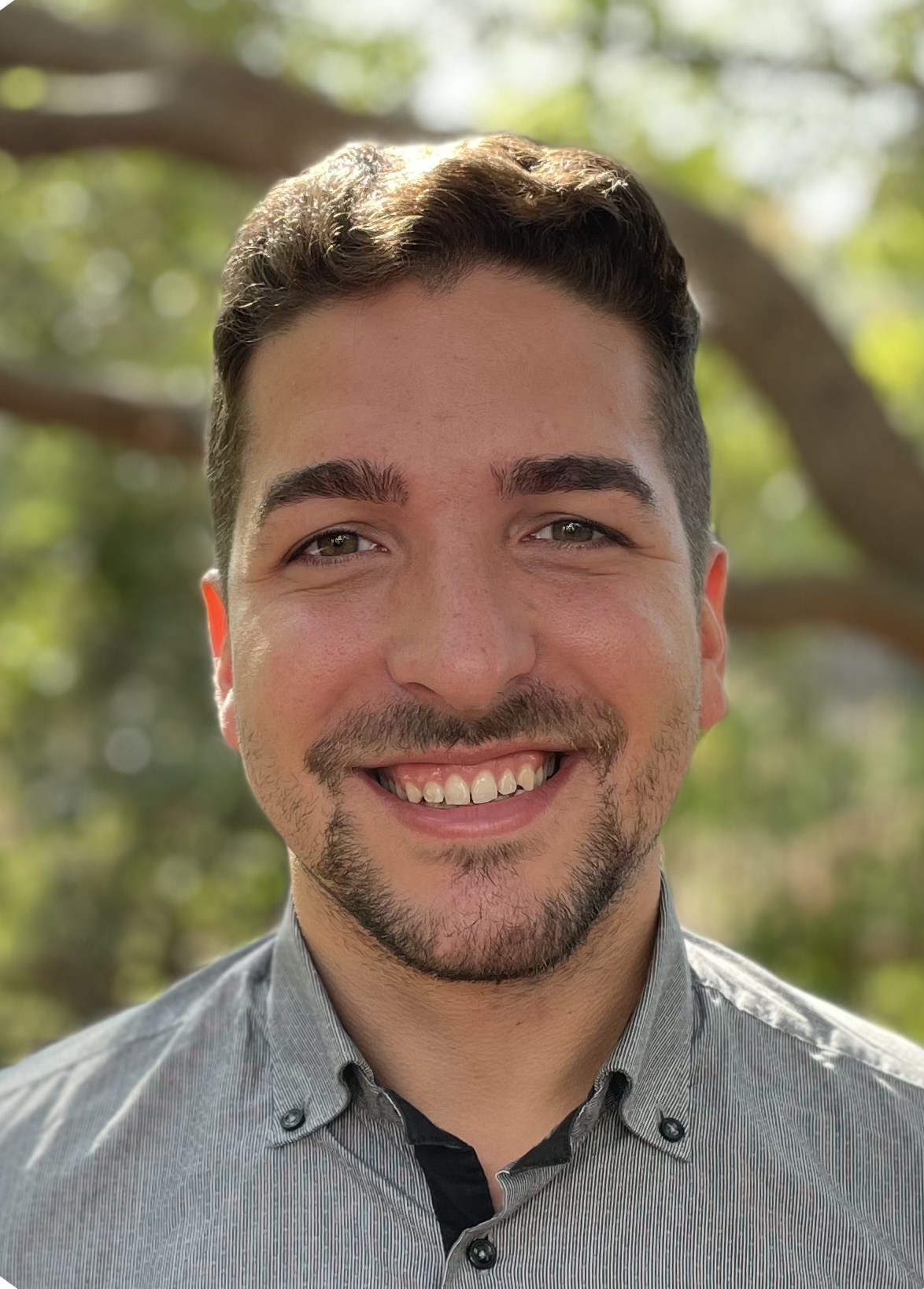}}]{Davide Caputo} obtained  his  Ph.D.  in  Computer Science at  the  University  of  Genova (Italy) in 2022 under the supervision of Alessio Merlo  and Luca Verderame.  His research topic focuses on Mobile Security and IoT Security. Now he is working as Cybersecurity Engineer at Talos s.r.l.s.
\end{IEEEbiography}
\vspace{-1cm}
\begin{IEEEbiography}[{\includegraphics[width=1in,height=1.25in,clip,keepaspectratio]{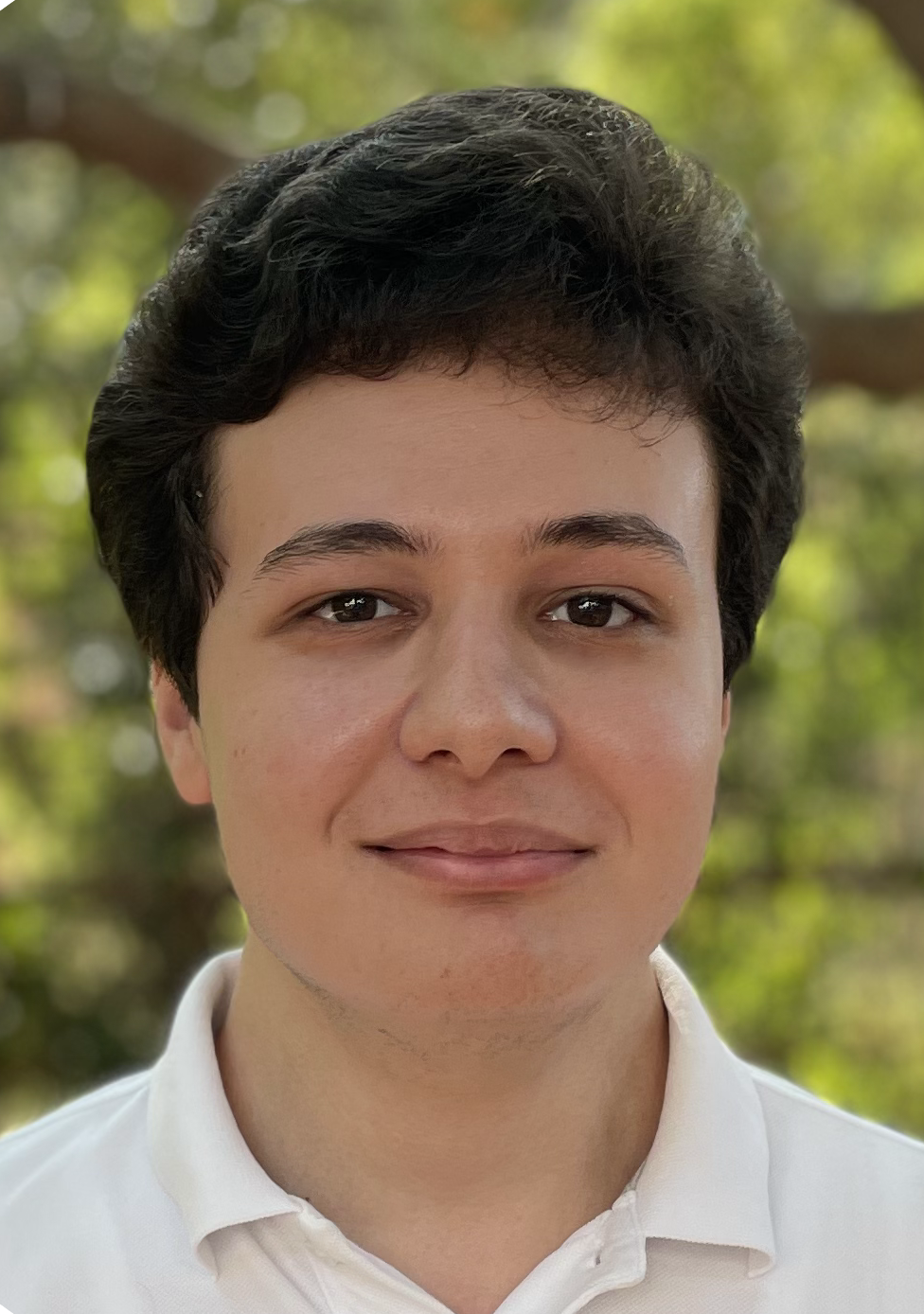}}]{Francesco Pagano} obtained both his BSc and MSc in Computer Engineering at the University of Genova. His research topic focuses on Mobile Security and IoT Security. Currently, he is a Ph.D. student at the University of Genova within the Computer Security Laboratory (CSecLab), under the supervision of Alessio Merlo and Luca Verderame.
\end{IEEEbiography}

\vspace{-1cm}
\begin{IEEEbiography}[{\includegraphics[width=1in,height=1.25in,clip,keepaspectratio]{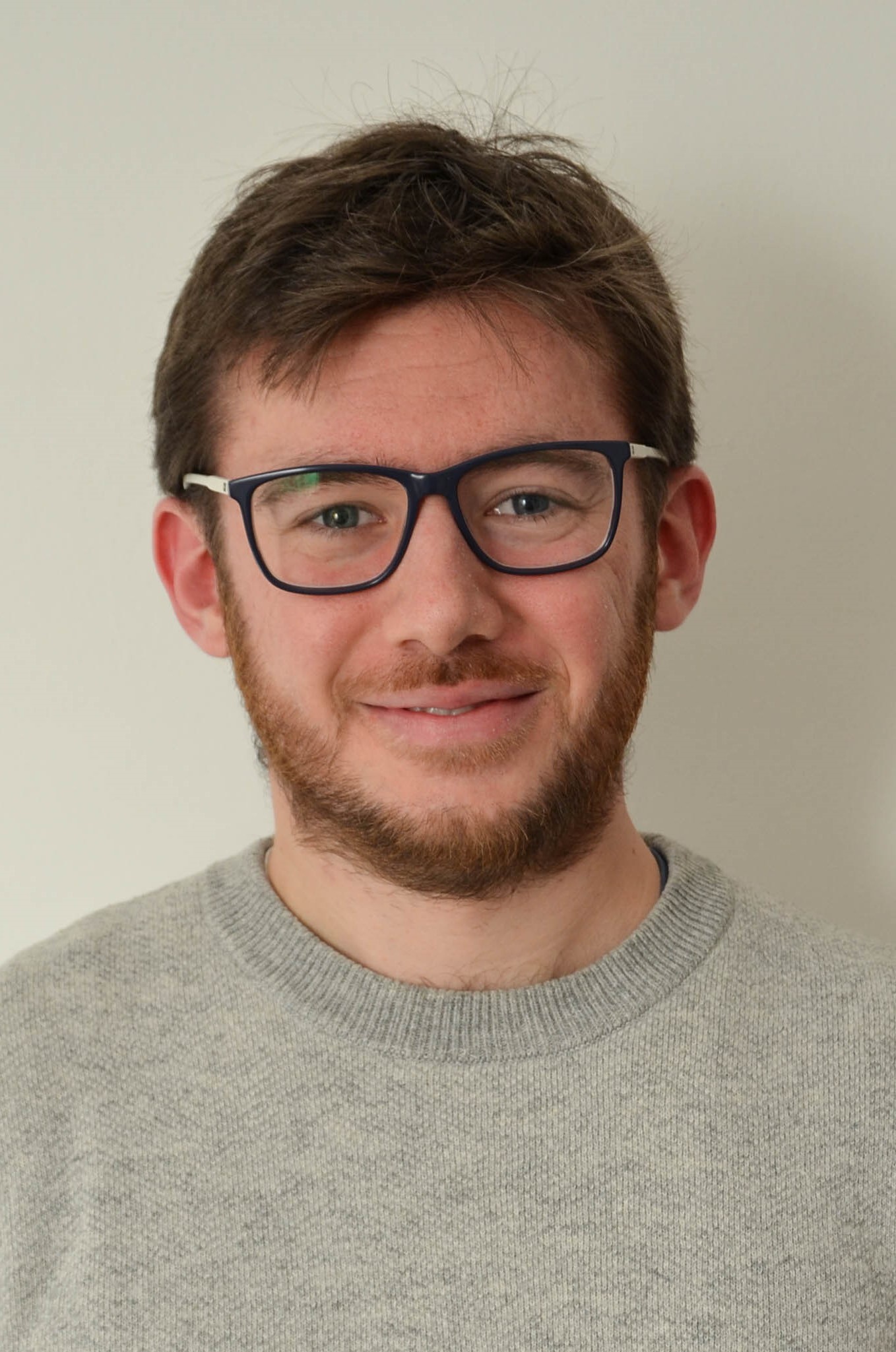}}]{Giovanni Bottino} obtained both his BSc and MSc in Computer Engineering at the University of Genova. Together with his colleague Francesco Pagano and under the supervision of Prof. Alessio Merlo and Davide Caputo, he focused his master's thesis on the issues of Mobile Privacy with the development of the HideDroid app. Actually he is working as a software developer at the RINA S.p.A. company in Genova.
\end{IEEEbiography}
\vspace{-1cm}
\begin{IEEEbiography}[{\includegraphics[width=1in,height=1.25in,clip,keepaspectratio]{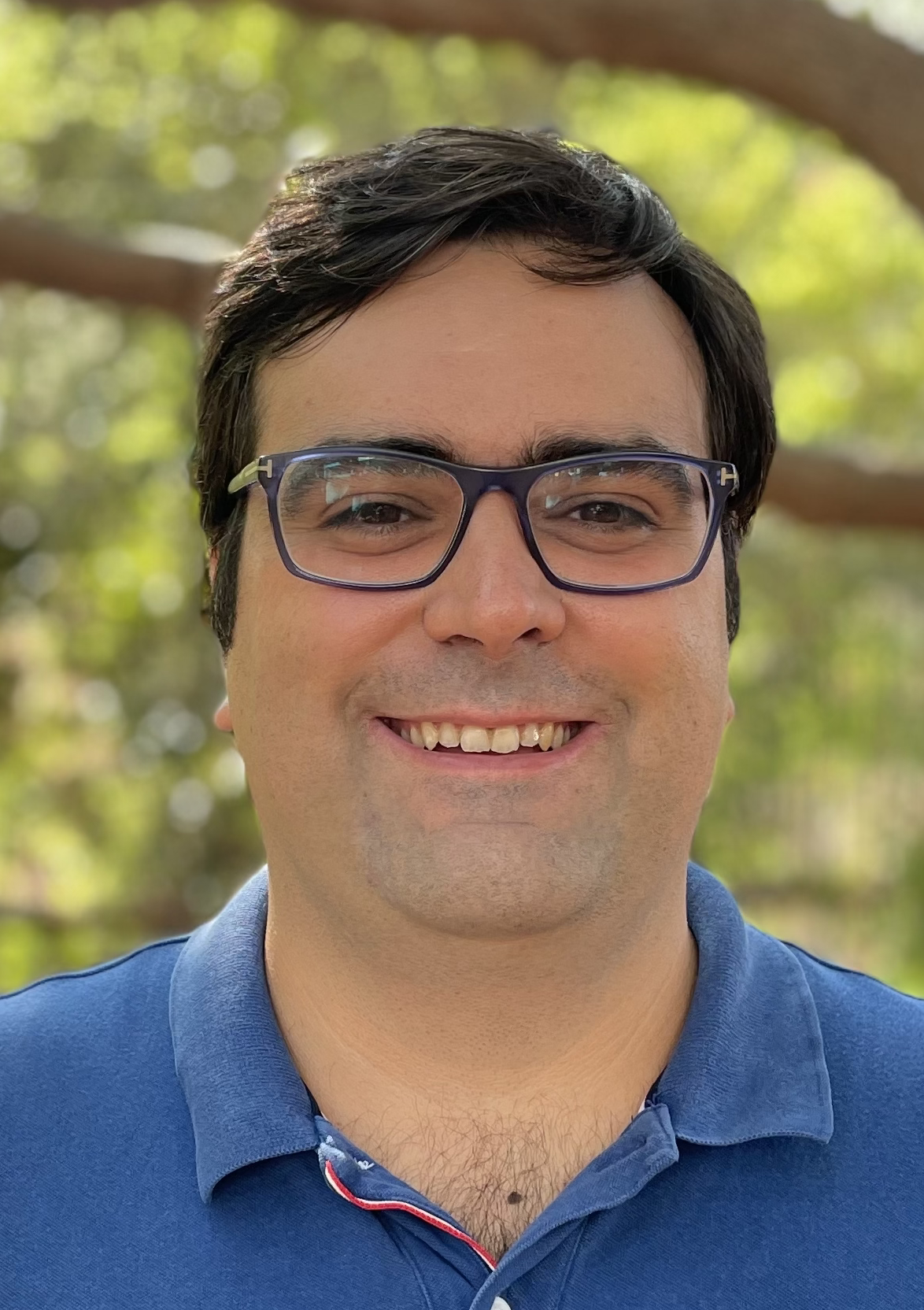}}]{Luca Verderame} is an Assistant Professor in Computer Engineering at the University of Genoa (Italy).
He obtained his PhD in Electronic, Information, Robotics, and Telecommunication Engineering in 2016, where he worked on mobile security. His research interests mainly cover the security of application ecosystems. Luca is also the CEO and Co-founder of Talos s.r.l.s., a cybersecurity SME and university spin-off. 
\end{IEEEbiography}

\vspace{-1cm}
\begin{IEEEbiography}[{\includegraphics[width=1in,height=1.25in,clip,keepaspectratio]{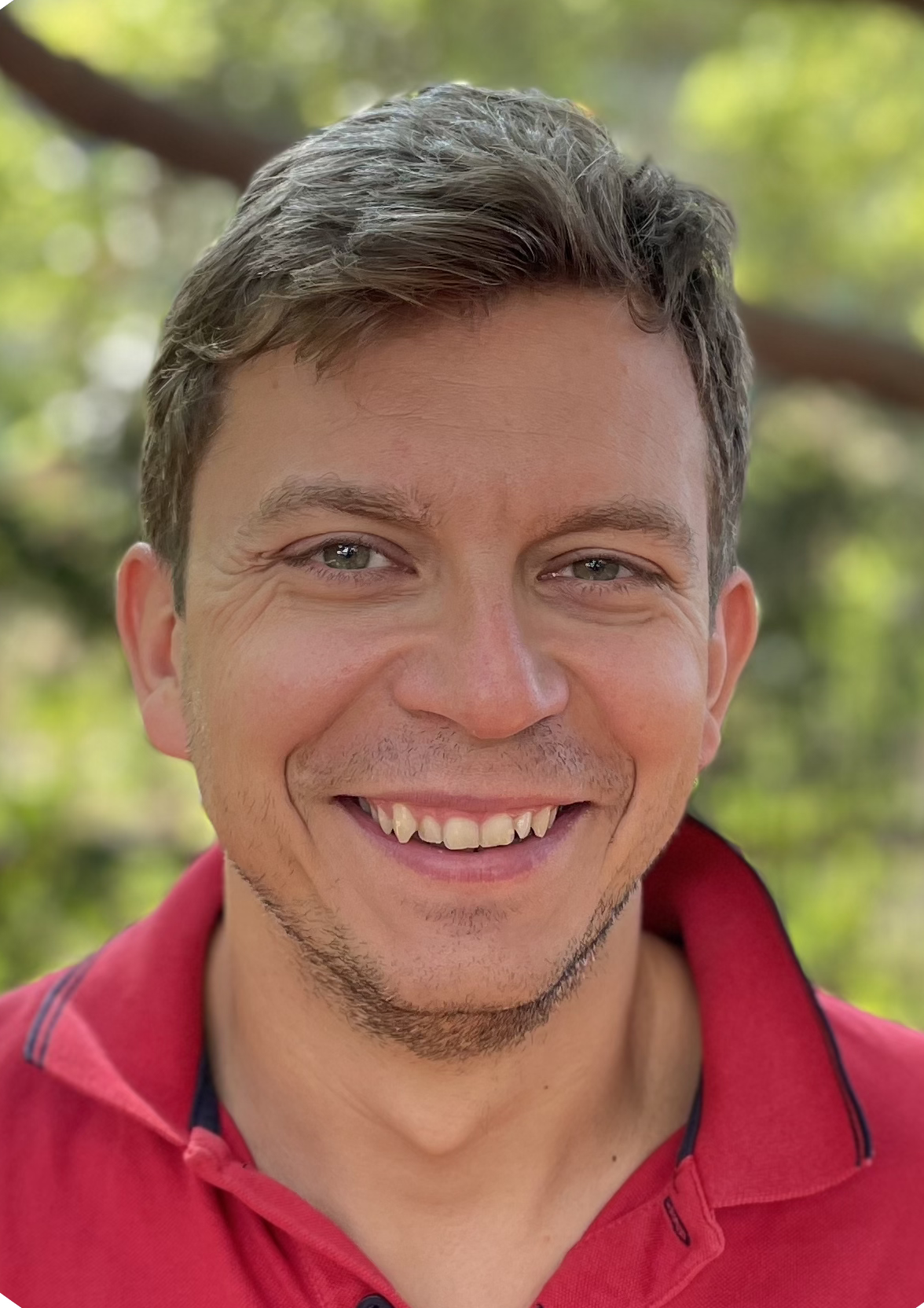}}]{Alessio Merlo} is an Associate Professor in Computer Engineering at the University of Genova where he leads the Mobile Security research group.  His main research interests focus on Mobile and IoT Security. He published more than 100 scientific papers in international conferences and journals. 
\end{IEEEbiography}




\end{document}